\documentclass[Afour,sageh,times]{sagej}

\usepackage{moreverb,url}

\usepackage[colorlinks,bookmarksopen,bookmarksnumbered,citecolor=red,urlcolor=red]{hyperref}

\newcommand\BibTeX{{\rmfamily B\kern-.05em \textsc{i\kern-.025em b}\kern-.08em
T\kern-.1667em\lower.7ex\hbox{E}\kern-.125emX}}

\def\volumeyear{2019}
\usepackage{makeidx}         
\usepackage{graphicx}        
\usepackage{float}
\usepackage{enumerate}
\usepackage{subfig}
\usepackage{sidecap}
\usepackage{amssymb}
\usepackage{amsmath}
\usepackage{bbm}
\usepackage{mathtools}
\usepackage{boxedminipage}

\newtheorem{theorem}{Theorem}
\newtheorem{example}{Example}
\newtheorem{definition}{Definition}

\newcommand{\revised}[1]{{\leavevmode\color{black}#1}}

\begin{document}

\runninghead{Karen Leung, et al.}

\title{Backpropagation through Signal Temporal Logic Specifications: Infusing Logical Structure into Gradient-Based Methods}

\author{
Karen Leung\affilnum{1},
Nikos Ar\'echiga\affilnum{2},
Marco Pavone\affilnum{1}}

\affiliation{\affilnum{1}Stanford University, Stanford, CA 94305, USA\\
\affilnum{2}Toyota Research Institute, Los Altos, CA 94022, USA}
\corrauth{Marco Pavone, 496 Lomita Mall, Rm. 261 Stanford, CA 94305
\email{pavone@stanford.edu}}

\begin{abstract}
This paper presents a technique, named \texttt{stlcg}, to compute the quantitative semantics of Signal Temporal Logic (STL) formulas using computation graphs. \texttt{stlcg} provides a platform which enables the incorporation of logical specifications into robotics problems that benefit from gradient-based solutions.
Specifically, STL is a powerful and expressive formal language that can specify spatial and temporal properties of signals generated by both continuous and hybrid systems.
The quantitative semantics of STL provide a robustness metric, i.e., how much a signal satisfies or violates an STL specification.
In this work, we devise a systematic methodology for translating STL robustness formulas into computation graphs.
With this representation, and by leveraging off-the-shelf automatic differentiation tools, we are able to efficiently backpropagate through STL robustness formulas and hence enable a natural and easy-to-use integration of STL specifications with many gradient-based approaches used in robotics.
Through a number of examples stemming from various robotics applications, we demonstrate that \texttt{stlcg} is versatile, computationally efficient, and capable of incorporating human-domain knowledge into the problem formulation.
\end{abstract}


\maketitle


\section{Introduction}
\label{sec:intro}
There are rules or spatio-temporal constraints that govern how a system should operate. Depending on the application, a system could refer to a robotic system (e.g., autonomous car or drone), or components within an autonomy stack (e.g., controller, monitor, or predictor).
Such rules or constraints can be explicitly known based on the problem definition or stem from human domain knowledge. For instance, road rules dictate how drivers on the road should behave, or search-and-rescue protocols prescribe how drones should survey multiple regions of a forest.
Knowledge of such rules is not only critical for successful deployment and is also a useful form of inductive bias when designing such a system.

A common approach to translating rules or spatio-temporal constraints expressed as natural language into a mathematical representation is to use a logic-based formal language, a mathematical language that consists of logical operators and a grammar describing how to systematically combine logical operators in order to build more complex logical expressions.
While the choice of a logic language can be tailored towards the application, one of the most common logic languages used in robotics is Linear Temporal Logic (LTL) \citep{Pnueli1977,BaierKatoen2008,Kress-GazitLahijanianEtAl2018,WongpiromsarnTopcuEtAl2011,FainekosGirardEtAl2008}, a temporal logic language that can describe and specify temporal properties of a sequence of \emph{discrete} states (e.g., the traffic light will always eventually turn green). For example, \cite{FinucaneJingEtAl2010} uses LTL to synthesize a ``fire-fighting'' robot planner that patrols a ``burning'' house to rescue humans and remove hazardous items.
Part of the popularity of LTL is rooted in the fact that many planning problems with LTL constraints, if formulated in a certain way, can be cast into an automaton for which there are well-studied and tractable methods to synthesize a planner satisfying the LTL specifications \citep{BaierKatoen2008}.
However, the discrete nature of LTL makes its usage limited to high-level planners and faces scalability issues when considering high-dimensional, continuous, or nonlinear systems.
Importantly, LTL is incompatible with gradient-based methods since LTL operates on \emph{discrete} states.

Recently, Signal Temporal Logic (STL) \citep{MalerNickovic2004} was developed and, in contrast to LTL, STL is a temporal logic language that is specified over \emph{dense-time real-valued} signals, such as state trajectories produced from \emph{continuous} dynamical systems.
Furthermore, STL is equipped with \emph{quantitative semantics} which provide the \emph{robustness} value of an STL specification for a given signal, i.e., a continuous real-value that measures the degree of satisfaction or violation.
By leveraging the quantitative semantics of STL, it is possible to use STL within a range of gradient-based methods.
Naturally, STL is becoming increasingly popular in many fields including machine learning and deep learning \citep{Vazquez-ChanlatteDeshmukhEtAl2017,MaGaoEtAl2020}, reinforcement learning \citep{PuranicDeshmukhEtAl2020}, and planning and control \citep{RamanDonzeEtAl2014,MehrSadighEtAl2017,YaghoubiFainekos2019,LiuMehdipourEtAl2021}.
While STL is attractive in many ways, there are still algorithmic and computational challenges when taking into account STL considerations.
Unlike LTL, STL is not equipped with such automaton theory and thus the incorporation of STL specifications into the problem formulation may require completely new algorithmic approaches or added computational complexity.

The goal of this work is to develop a framework that enables STL to be easily incorporated in a range of applications---we view STL as a useful mathematical language and tool to provide existing techniques with added logical structure or logic-based inductive bias, but at the same time, \revised{we want to avoid, if possible, causing} significant changes to the algorithmic and computational underpinnings when solving the problem.
We are motivated by the wide applicability, vast adoption, computational efficiency, ease-of-use, and accessibility of modern automatic differentiation software packages that are prevalent in many popular programming languages.
Additionally, many of these automatic differentiation tools are utilized by popular deep learning libraries such as PyTorch \citep{PaszkeGrossEtAl2017} which are used widely throughout the learning, robotics, and controls communities and beyond.
To this end, we propose \texttt{stlcg}, a technique that translates any STL robustness formula into \revised{computation graphs which are the underlying structure behind automatic differentiation}.
By leveraging readily-available automatic differentiation tools, we are able to efficiently backpropagate through STL robustness formulas and therefore make STL amenable to gradient-based computations.
Furthermore, by specifically utilizing deep learning libraries that have in-built automatic differentiation functionalities, we are essentially casting STL and neural networks in the same computational language and thereby bridging the gap between spatio-temporal logic and deep learning.

\noindent {\bf \emph{Statement of contributions: }}The contributions of this paper are threefold. First, we describe the mechanism behind \texttt{stlcg} by detailing the construction of the computation graph for an arbitrary STL formula. Through this construction, we prove the correctness of \texttt{stlcg}, and show that it scales at most quadratically with input length and linearly in formula size. The translation of STL formalisms into computation graphs allows \texttt{stlcg} to inherit many benefits such as the ability to backpropagate through the graph to obtain gradient information; this abstraction also provides computational benefits such as enabling portability between hardware backends (e.g., CPU and GPU) and batching for scalability.
Second, we open-source our PyTorch implementation of \texttt{stlcg}; it is a toolbox that makes \texttt{stlcg} easy to combine with many existing frameworks and deep learning models.
Third, we highlight key advantages of \texttt{stlcg} by investigating a number of diverse example applications in robotics such as motion planning, logic-based parameter fitting, regression, intent prediction, and latent space modeling. We emphasize \texttt{stlcg}'s computational efficiency and natural ability to embed logical specifications into the problem formulation to make the output more robust and interpretable.

A preliminary version of this work appeared at the 2020 Workshop on Algorithmic Foundations of Robotics (WAFR). In this revised and extended version, we provide the following additional contributions: \revised{(i) A more detailed and pedagogical introduction to STL with additional illustrative examples, (ii) a more thorough description of the construction of all the STL computation graphs including details about the Until operator, (iii) an additional motion planning example demonstrating the properties of the Until operator, (iv) additional experiments for the parametric STL example, and (v) an example demonstrating the usage of \texttt{stlcg} in latent space models.}

\emph{Organization: }
In Section \ref{sec:prelim}, we cover the basic definitions and properties of STL, and
describe the graphical structure of STL formulas. Equipped with the graphical structure of an STL formula, Section \ref{sec:computation graph} draws connections to computation graphs and provides details on the technique that underpins \texttt{stlcg}.
We showcase the benefits of \texttt{stlcg} through a variety of case studies in Section~\ref{sec:examples}. We finally conclude in Section~\ref{sec:conclusion and future work} and propose exciting future research directions for this work.


\section{Preliminaries}
\label{sec:prelim}
In this section, we provide the definitions and syntax for STL, and describe the underlying graphical structure of the computation graph used to evaluate STL robustness formulas.

\subsection{Signals}\label{subsec:prelim signals}
\revised{
STL formulas are interpreted over \emph{signals}, a sequence of numbers representing real-valued, discrete-time outputs (i.e., continuous-time outputs sampled at finite time intervals) from any system of interest, such as a sequence of robot states, the temperature of a building, or the speed of a vehicle. In this work, we assume that the signal is sampled at uniform time steps $\Delta t$.
}

\revised{\begin{definition}[Signal]
A signal 
\[s_{t_i}^{t_T} = (x_i, t_i), \, (x_{i+1}, t_{i+1}), \ldots , (x_{T}, t_{T})\]
is an ordered ($t_{j-1} < t_j$) finite sequence of states $x_j\in\mathbb{R}^n$ and their associated times $t_j\in\mathbb{R}$ starting from time $t_i$ and ending at time $t_T$.
\label{def:signal}
\end{definition}

Given a signal, a subsignal is a contiguous fragment of a signal. 

\begin{definition}[Subsignal]
Given a signal $s_{t_0}^{t_T}$, a subsignal of $s_{t_0}^{t_T}$ is a signal $s_{t_i}^{t_T}$ where $i \geq 0$ and $t_i \geq t_0$.
\end{definition}}

\noindent \revised{\emph{Remark:} For notational brevity, when the final time of a signal or subsignal is clear from context or abstractly fixed, we drop the superscript and write $s_{t}$ to represent the signal starting from time $t$ until the end of the signal.
Similarly, we drop both the subscript and superscript and write $s$ to represent a signal whenever the start and end times are clear from context, or abstractly fixed.}

\subsection{Signal Temporal Logic: Syntax and Semantics}
\label{subsec:prelim stl syntax and semantics}

STL formulas are defined recursively according to the following grammar \citep{BartocciDeshmukhEtAl2018,MalerNickovic2004},
\begin{align}
\phi::= &~~\top ~|~ \mu_c ~|~ \neg\phi ~|~ \phi \wedge \psi ~|~  \phi\,\mathcal{U}_{[a,b]}\,\psi.
\label{eq:STL grammar}
\end{align}
Like in many other STL (and LTL) literature, the STL grammar presented in \eqref{eq:STL grammar} is written in Backus-Naur form, a context-free grammar commonly used by developers of programming languages to specify the syntax rules of a language. The grammar describes a set of rules that outline the syntax of the language and ways to construct a ``string'' (i.e., an STL formula).
An STL formula $\phi$ is generated by selecting an expression from the list of expressions on the right separated by the pipe symbol ($~|~$) in a recursive fashion.
Now, we describe each element of the grammar, \revised{and the semantics of each STL operator thereafter}:
\begin{itemize}
\item $\top$ means true and we can set an STL formula as $\phi = \top$
\item $\mu_c$ is a predicate of the form $\mu(x) > c$ where $c\in\mathbb{R}$ and $\mu: \mathbb{R}^n \rightarrow \mathbb{R}$ is a differentiable function that maps the state $x\in\mathbb{R}^n$ to a scalar value. An STL formula can be defined as $\phi = \mu(x) > c$. \revised{Predicates form the building blocks of an STL formula.}
\item $\neg \phi$ denotes applying the \emph{Not} operation $\neg$ onto an STL formula $\phi$. \revised{E.g., let $\phi = \mu(x) < c$, then $\neg \phi$ is also an STL formula.}
\item $\phi \wedge \psi$ denotes combining $\phi$ with another STL formula $\psi$ via the \emph{And} operation $\wedge$. \revised{E.g., let $\phi$ and $\psi$ be two STL formulas. Then $\phi \wedge \psi$ is also an STL formula.}
\item $\phi\,\mathcal{U}_{[a,b]}\, \psi$ denotes combining $\phi$ with another STL formula $\psi$ via the \emph{Until} temporal operation $\mathcal{U}_{[a,b]}$. \revised{The interval $[a,b]\subseteq [0, t_T]$ is a time interval which the Until operator performs on. \revised{E.g., let $\phi$ and $\psi$ be two STL formulas. Then $\phi \,\mathcal{U}_{[a,b]}\, \psi$ is also an STL formula.}}
\end{itemize}
One can generate an arbitrarily complex STL formula by recursively applying STL operations. See Example \ref{eg:stl construction} for a concrete example.
Additionally, other commonly used logical connectives and temporal operators can be derived using the grammar presented in \eqref{eq:STL grammar},
\begin{flalign*}
\qquad \phi \vee \psi &= \neg(\neg\phi \wedge \neg\psi) & & \text{\textit{(Or)}} \quad\\
\qquad \phi \Rightarrow \psi &= \neg\phi \vee \psi & & \text{\textit{(Implies)}} \quad\\
\qquad \lozenge_{[a,b]}\, \phi &= \top \,\mathcal{U}_{[a,b]}\,\phi & & \text{\textit{(Eventually)}} \quad\\
\qquad \square_{[a,b]}\,\phi &= \neg \lozenge_{[a,b]} (\neg \phi) & & \text{\textit{(Always)}}.\quad
\end{flalign*}
\revised{\emph{Remark:} When the time interval $[a,b]$ is dropped in the temporal operators, it corresponds to $[a,b] = [0,t_T]$ where $t_T$ is the final time, i.e., the time interval is over the entire signal.
For instance, $\square\, \phi$ refers to $\phi$ being true for \emph{all} time steps over the signal, while $\square_{[a,b]} \phi$ means that $\phi$ is only true within the time interval $[a,b]$.}
\begin{example}\label{eg:stl construction}
Let $x\in\mathbb{R}^n$, $\phi_1 = \mu_1(x) < c_1$ and $\phi_2 = \mu_2(x) \geq c_2$.\footnote{Equality and the other inequality relations can be derived from the STL grammar in \eqref{eq:STL grammar}, i.e., $\mu(x) < c \Leftrightarrow -\mu(x) > -c$, and $\mu(x) = c \Leftrightarrow    \neg(\mu(x) < c) \wedge \neg(\mu(x) > c)$.} Then, for a given signal $s_t$, the formula $\phi_3 = \phi_1 \wedge \phi_2$ is true if both $\phi_1$ and $\phi_2$ are true. Similarly, $\phi_4 = \square_{[a,b]}\phi_3$ is true if $\phi_3$ is always true over the entire interval $[t+a, t+b]$.
\end{example}

\noindent
\revised{Given a signal $s_t$ starting at time $t$ with arbitrary (finite) length,} we use the notation \(s_t \models \phi\) to denote that a signal \(s_t\) satisfies an STL formula \(\phi\)
according to the formal semantics below.

\revised{We first informally describe the temporal operators}:
\begin{itemize}
    \item $s_t \models \phi \, \mathcal{U}_{[a,b]}\, \psi$ if there is a time $t^\prime\in[t+a,t+b]$ such that $\phi$ holds for all time before and including $t^\prime$ and $\psi$ holds at time $t^\prime$.
    \item $s_t\models \lozenge_{[a,b]} \, \phi$ if at some time $t\in[t+a,t+b]$, $\phi$ holds at least once.
    \item $s_t\models \square_{[a,b]}\,\phi$ if $\phi$ holds for all $t \in [t+a,t+b]$.
\end{itemize}

Formally, the \emph{Boolean} semantics of a formula (i.e., formulas to determine if the formula is true or false) with respect to a signal $s_t$ are defined as follows.

\begin{boxedminipage}{0.45\textwidth}
\begin{center}
\textbf{Boolean Semantics}
\end{center}
\vspace{-5mm}
\begin{align*}
    s_t & \models \mu_c &\Leftrightarrow  \quad& \; \mu(x_t) > c\\
    s_t & \models \neg \phi  &\Leftrightarrow  \quad& \; \neg(s_t \models \phi)\\
    s_t & \models \phi \wedge \psi & \Leftrightarrow  \quad& \; (s_t \models \phi) \wedge (s_t \models \psi)\\
    s_t & \models \phi \vee \psi & \Leftrightarrow  \quad& \; (s_t \models \phi) \vee (s_t \models \psi)\\
    s_t & \models \phi \Rightarrow \psi & \Leftrightarrow  \quad& \; \neg(s_t \models \phi) \vee (s_t \models \psi)\\
    s_t & \models \lozenge_{[a,b]} \phi & \Leftrightarrow  \quad& \; \exists t^\prime \in [t+a, t+b] \;\;\mathrm{s.t.} \; s_{t^\prime} \models \phi\\
    s_t & \models \square_{[a,b]} \phi & \Leftrightarrow  \quad& \; \forall t^\prime \in [t+a, t+b] \;\;s.t. \;  s_{t^\prime} \models \phi\\
    s_t & \models \phi \,\mathcal{U}_{[a,b]}\, \psi & \Leftrightarrow  \quad& \; \exists t^\prime \in [t+a, t+b] \;\;s.t. \; (s_{t^\prime} \models \psi)\; \\
    &&& \quad \qquad  \wedge \;(\forall \tau \in [t, t^\prime], \, s_\tau \models  \phi)
\end{align*}
\end{boxedminipage}

\vspace{3mm}

Furthermore, STL admits a notion of \emph{robustness}, that is, it has \emph{quantitative semantics} which calculate how much a signal satisfies or violates a formula. Positive robustness values indicate satisfaction, while negative robustness values indicate violation. Like Boolean semantics, the quantitative semantics of a formula with respect to a signal $s_t$ is defined as follows,
\vspace{3mm}

\begin{boxedminipage}{0.49\textwidth}
\begin{center}
\textbf{Quantitative Semantics}
\end{center}
\vspace{-2mm}
{\small
\begin{equation}
\begin{aligned}
\rho(s_t, \top) & \:= \: \rho_{\max} \qquad \text{where $\rho_{max} > 0$}\\
\rho(s_t, \mu_c) & \:= \: \mu(x_t) - c\\
\rho(s_t, \neg \phi) & \:= \: -\rho(s_t,\phi)\\
\rho(s_t, \phi \wedge \psi) & \:= \: \min(\rho(s_t,\phi), \rho(s_t,\psi))\\
\rho(s_t, \phi \vee \psi) & \:= \: \max(\rho(s_t,\phi), \rho(s_t,\psi))\\
\rho(s_t, \phi \Rightarrow \psi) & \:= \: \max(-\rho(s_t,\phi), \rho(s_t,\psi))\\
\rho(s_t, \lozenge_{[a,b]} \phi) & \:= \: \max_{t^\prime \in [t+a, t+b]}\rho(s_{t^\prime},\phi)\\
\rho(s_t, \square_{[a,b]} \phi) & \:= \: \min_{t^\prime \in [t+a, t+b]}\rho(s_{t^\prime},\phi)\\
\rho(s_t, \phi \,\mathcal{U}_{[a,b]}\psi) & = \max_{t^\prime \in [t+a, t+b]} \min\lbrace\rho(s_{t^\prime}, \psi), \min_{\tau\in [t, t^\prime]} \rho(s_{\tau}, \phi)\rbrace.
\end{aligned}
\label{eq:STL robustness}
\end{equation}
}
\end{boxedminipage}
\vspace{3mm}

\begin{figure*}[t]
    \centering
    \includegraphics[width=\textwidth]{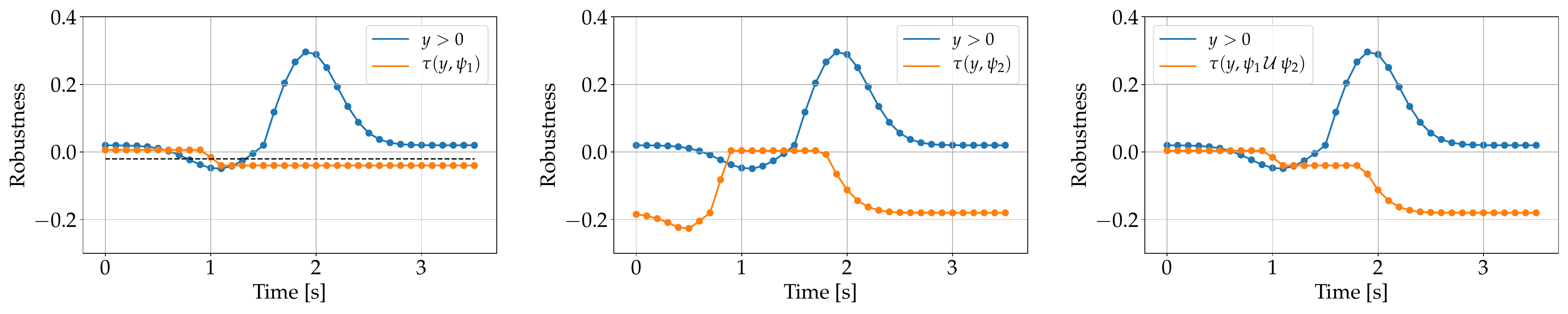}
    \caption{\revised{Robustness traces of $\psi_1 = \lozenge\, \square_{[0,0.4]} \, ( y < -0.02 )$ (left), $\psi_2 = \lozenge_{[0, 0.8]}\, \square_{[0, 0.4]} \, ( y > 0.2 )$ (center), and $\psi_1\, \mathcal{U}\, \psi_2$ (right). The input signal $y$ is shown in blue, and the corresponding robustness trace is shown in orange.}}
    \label{fig:until toy example}
\end{figure*}

\revised{We refer to the quantitative semantics as \emph{robustness formulas}. The formulas represent a natural way to measure the degree of satisfaction or violation, and we highlight a few formulas to illustrate the intuition behind the robustness formulas. 
The robustness formula for the predicate $\mu(x) > 0$ measures how much greater $\mu(x)$ is than $c$.
For the And operation, $\phi \wedge \psi$, since we want $\phi$ and $\psi$ to both hold, the robustness formula computes the lower of the two robustness values. Similarly, for the Or operation, we want either $\phi$ or $\psi$ to hold, therefore we consider the higher of the two robustness values. 
For the Until operator, the $\min_{\tau \in [t, t^\prime]} \rho(s_\tau, \phi)$ term corresponds to the condition that $\phi$ needs to hold for all times in $[t,t^\prime]$ where $t^\prime \in [t+a, t+b]$. Then the outer $\min$ operation corresponds to the condition that $\psi$ also needs to hold at time $t^\prime$. Finally, the outer $\max$ operation requires that the previous two conditions holds at least once within the time interval $[t+a, t+b]$.
}

Further, we define the \emph{robustness trace} as a sequence of robustness values of every subsignal $s_{t_i}$ of signal $s_{t_0}$, $t_i \geq t_0$. \revised{The definition of robustness trace is presented formally below, and 
Example~\ref{eg:robustness trace} visualizes the robustness traces of various temporal operators.}

\revised{\begin{definition}[Robustness trace]
Given a signal $s_{t_0}$ and an STL formula $\phi$,  the robustness trace $\tau(s_{t_0},\phi)$ is a sequence of robustness values of $\phi$ for each subsignal of $s_{t_0}$. Specifically,
\[\tau(s_{t_0}, \phi) = \rho(s_{t_0}, \phi),\,\rho(s_{t_1}, \phi),\ldots, \rho(s_{t_T}, \phi)\]
where $t_{i-1} < t_i$.
\end{definition}}

\revised{
\begin{example}\label{eg:robustness trace}
Let $\psi_1 = \lozenge\, \square_{[0,0.4]} \, ( y < -0.02 )$ which specifies that a signal $y$ should, at some time in the future, stay below -0.02 for 0.4 seconds. Let $\psi_2 = \lozenge_{[0, 0.8]}\, \square_{[0, 0.4]} \, ( y > 0.2 )$ which specifies that a signal $y$ should, within 0.8 seconds in the future, be greater than 0.2 for 0.4 seconds. Then a formula $\psi_1 \, \mathcal{U} \, \psi_2$ specifies that $\psi_1$ must always hold before $\psi_2$ holds.  
Figure~\ref{fig:until toy example} shows the signal $y$ (blue) and the corresponding robustness trace (orange) for each of the formulas. The robustness value $\rho(y, \psi_1\, \mathcal{U}\,\psi_2)$ is the robustness value evaluated at time zero on the plot on the right.
\end{example}
}

\subsection{Graphical Structure of STL Formulas}\label{subsec:prelim graphical properties}
Recall that STL formulas are defined recursively according to the grammar in \eqref{eq:STL grammar}.
We can represent the recursive structure of an STL formula with a parse tree $\mathcal{T}$ by identifying subformulas of each STL operation that was applied.

\begin{definition}[Subformula]
\label{def:subformula}
Given an STL formula $\phi$, a subformula of $\phi$ is a formula to which the outer-most (i.e., last) STL operator is applied to.
The operators $\wedge$, $\vee$, $\Rightarrow$, and $\mathcal{U}$ will have two corresponding subformulas, while predicates are not defined recursively and thus have no subformulas.
\end{definition}

\begin{definition}[STL Parse Tree]
\label{def:parse tree}
An STL parse tree  $\mathcal{T}$ represents the syntactic structure of an STL formula according to the STL grammar (defined in \eqref{eq:STL grammar}). Each node represents an STL operation and it is connected to its corresponding subformula(s).

\end{definition}

For example, the subformula of $\square\,\phi$ is $\phi$, and the subformulas of $\phi \,\wedge \, \psi$ are $\phi$ and $\psi$.
Given an STL formula, the root node of its corresponding parse tree represents the outermost operator for the formula. This node is connected to the outermost operator of its subformula(s), and so forth.
Applying this recursion, each node of the parse tree represents each operation that makes up the STL formula; the outermost operator is the root node and the innermost operators are the leaf nodes (i.e., the leaf nodes represent predicates or the terminal True operator.).

\begin{figure*}[t]
    \centering
    \subfloat[Parse tree $\mathcal{T}$.]{
    \label{fig:parse tree}
    \includegraphics[width=20mm]{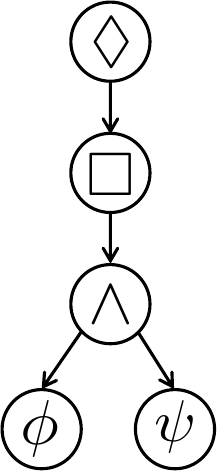}} $\qquad\qquad \qquad$ \subfloat[Computation graph $\mathcal{G}$.]{
    \label{fig:computation graph}
    \includegraphics[width=20mm]{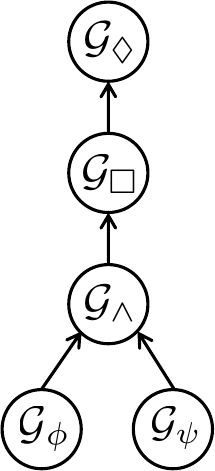}} $\qquad\qquad \qquad$\subfloat[Visualization of $\mathcal{G}$ using the \texttt{stlcg} toolbox.]{
    \label{fig:stlcg}
    \includegraphics[width=51mm]{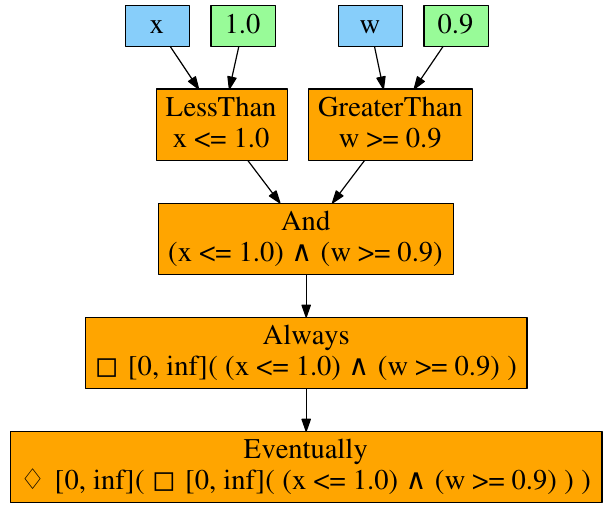}}
\caption{An illustration of a parse tree (left) and a computation graph (middle) for an STL formula $\lozenge \, \square \, (\phi \wedge \psi)$ where $\phi = w \geq 0.9$ and $\psi = x \leq 1.0$. On the right is a visualization generated from the \texttt{stlcg} toolbox. The blue nodes represent input signals ($x$ and $w$) into the computation graph, the green nodes represent parameters of the predicates, and the orange nodes represent STL operations.}
\label{fig:parse tree and computation graph example}
\end{figure*}

An example of a parse tree $\mathcal{T}_\theta$ for a formula $\theta = \lozenge \, \square \, (\phi \wedge \psi)$ is illustrated in Figure~\ref{fig:parse tree} where $\phi$ and $\psi$ are assumed to be predicates.
By flipping the direction of all the edges in $\mathcal{T}_\theta$, we obtain a directed acyclic graph $\mathcal{G}_\theta$, shown in Figure~\ref{fig:computation graph}.
At a high level, signals are passed through the root nodes, $\mathcal{G}_\phi$ and $\mathcal{G}_\psi$, to produce $\tau(s, \phi)$ and $\tau(s, \psi)$, robustness traces of $\phi$ and $\psi$. The robustness traces are then passed through the next node of the graph, $\mathcal{G}_\wedge$, to produce the robustness trace $\tau(s, \phi \wedge\psi)$, and so forth until the root node is reached and the output is $\tau(s, \lozenge \, \square \, (\phi \wedge \psi))$.

\section{STL Robustness Formulas as Computation Graphs}
\label{sec:computation graph}

We first describe how to represent each STL robustness formula described in \eqref{eq:STL robustness} as a computation graph, and then show how to combine them together to form the overall computation graph that computes the robustness trace of any given STL formula.
Further, we provide smooth approximations to the $\max$ and $\min$ operations to help make the gradients smoother when backpropagating through the graph, and introduce a new robustness formula which addresses some limitations of using the $\max$ and $\min$ functions.
The resulting computational framework, \texttt{stlcg}, is implemented using PyTorch \citep{PaszkeGrossEtAl2017} and  the code can be found at \url{https://github.com/StanfordASL/stlcg}. Further, this toolbox includes a graph visualizer, illustrated in Figure~\ref{fig:stlcg}, to show the graphical representation of the STL formula and how it depends on the inputs and parameters from the predicates.

\subsection{Computation Graphs and Backpropagation}
\begin{figure}[tb]
    \centering
    \includegraphics[width=0.3\textwidth]{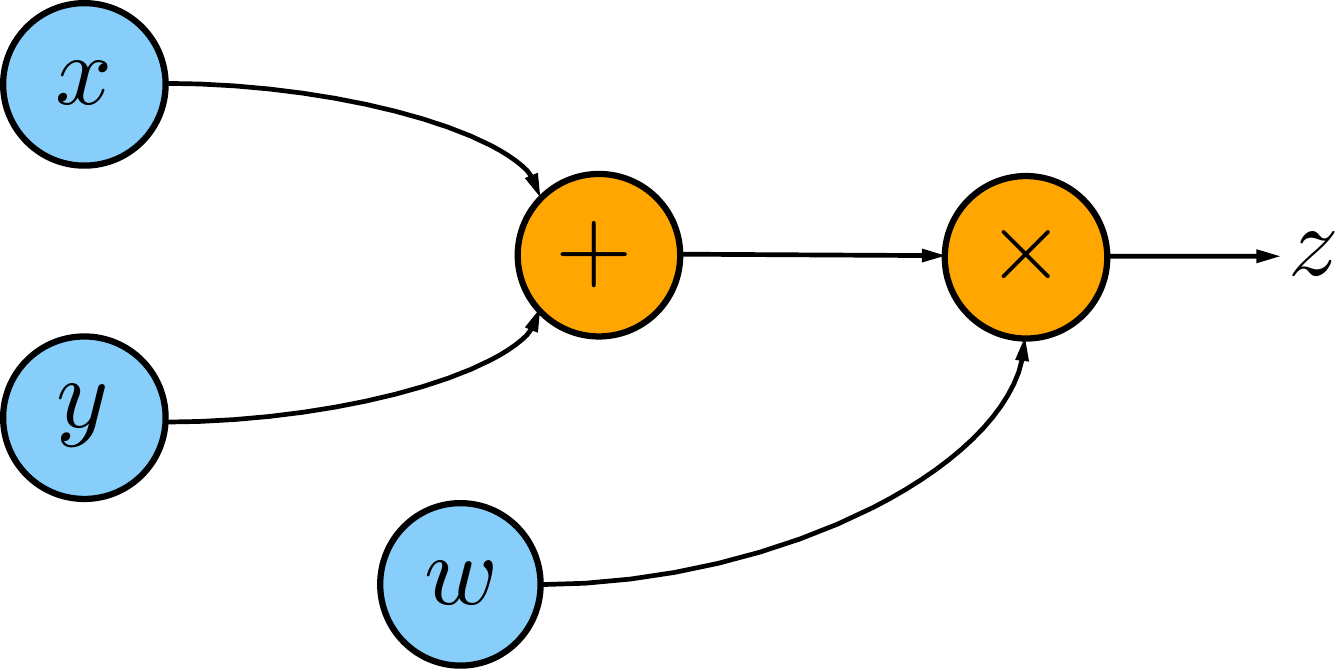}
    \caption{Computation graph of the mathematical operation $z=w(x+y)$.}
    \label{fig:computation graph example}
\end{figure}

A computation graph is a directed graph where the nodes correspond to operations or variables.
Variable values are passed through operation nodes, and the outputs can feed into other nodes and so forth.
For example, the computation graph of the mathematical operation $z = w(x + y)$ is illustrated in Figure~\ref{fig:computation graph example} where $x$, $y$, and $w$ are inputs into the graph, $+$ and $\times$ are mathematical operations, and $z$ is the output.

When the operations are differentiable, we can leverage the chain rule and therefore backpropagate through the computation graph to obtain derivatives of the output with respect to any variable that the output depends on.
\revised{
Recall, for a function that is composed of another other function $f(g(x))$, the chain rule ($\frac{\partial f}{\partial x} = \frac{\partial f}{\partial g} \frac{\partial g}{\partial x}$), describes how to compute the derivative of the function with respect to the input $x$. If the function is composed of $N$ functions, i.e., $f(g_N(...(g_2(g_1(x)))...))$, then using the chain rule, we have $\frac{\partial f}{\partial x} = \frac{\partial f}{\partial g_N} \left(\prod_{i=1}^{N-1} \frac{\partial g_{i+1}}{\partial g_i}\right)\frac{\partial g_1}{\partial x}$. This means for a function that is made up of a number of nested operations (e.g., a deep neural network or an STL robustness formula), we just need to know how to compute the derivative of each sub-operation and then multiply the derivatives together. When the function is expressed as a computation graph, computing the derivative of an output with respect to an input variable amounts to computing the derivative of all the edges that connect the two nodes together. 
Example~\ref{eg:backprop} illustrates how to perform backpropagation on a computation graph for a simple function $z = w(x + y)$.

\begin{example}\label{eg:backprop}
Consider the computation graph shown in Figure~\ref{fig:computation graph example}. In Figure~\ref{fig:backprop}, we have named each intermediate operation (orange nodes) with a variable $z^{(i)}, i=1,2$. The derivatives for each operation/edge are shown in red.
We have assigned values for the input variables $x=1, y=2, w=3$ and these values are shown in green. The values after each subsequent operation, $z^{(1)}=3, z^{(2)}=9$ are also shown in green. Using the chain rule, we have, 
\allowdisplaybreaks
\begin{align*}
\frac{\partial z}{\partial x} &= \frac{\partial z}{\partial z^{(2)}}\frac{\partial z^{(2)}}{\partial z^{(1)}}\frac{\partial z^{(1)}}{\partial x}\\
                              &= w\\
                              &= 3, \\
\frac{\partial z}{\partial y} &= \frac{\partial z}{\partial z^{(2)}}\frac{\partial z^{(2)}}{\partial z^{(1)}}\frac{\partial z^{(1)}}{\partial y}\\
                              &= w\\
                              &= 3, \\
\frac{\partial z}{\partial w} &= \frac{\partial z}{\partial z^{(2)}}\frac{\partial z^{(2)}}{\partial w}\\
                              & = z^{(1)}\\
                              & = x + y\\
                              & = 3.
\end{align*}
\end{example}

\begin{figure}[tb]
    \centering
    \includegraphics[width=0.45\textwidth]{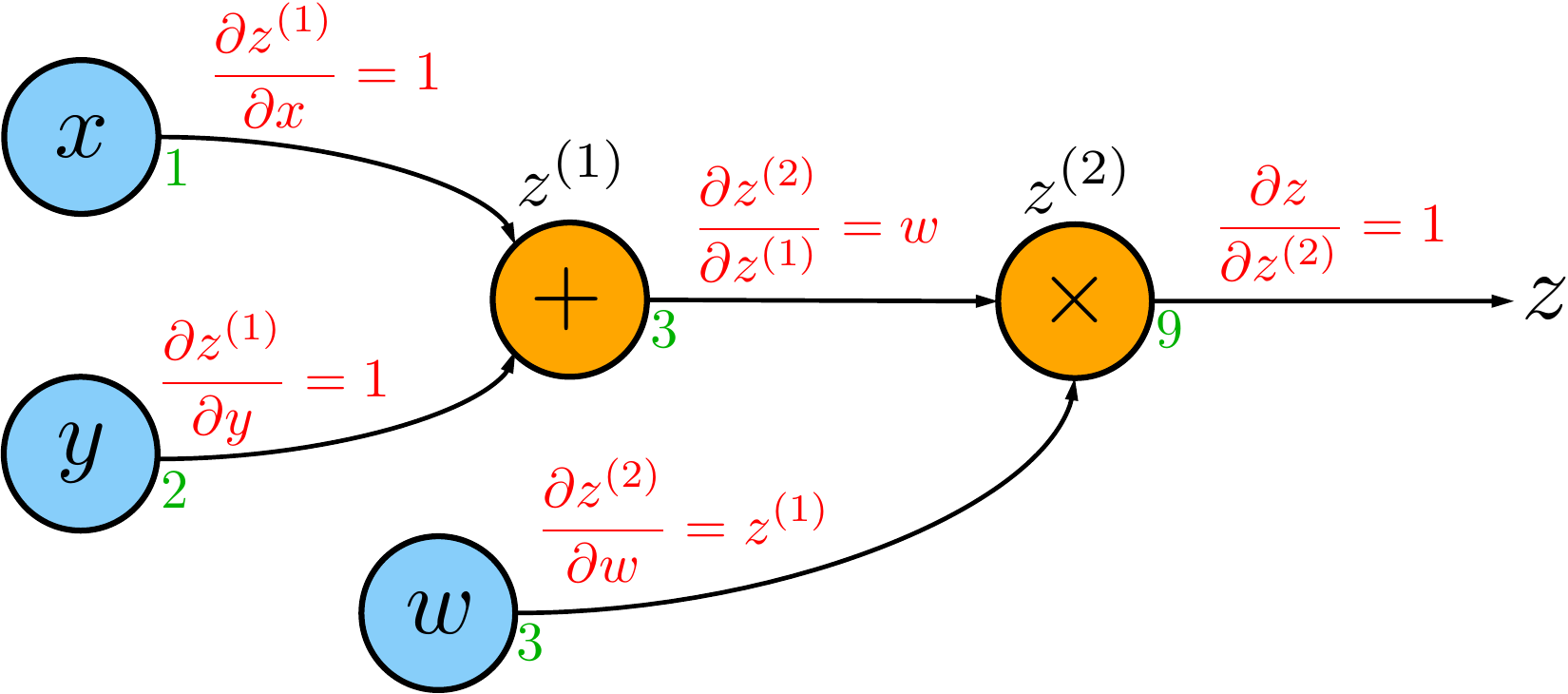}
    \caption{\revised{Backpropagation example with $z=w(x+y)$.}}
    \label{fig:backprop}
\end{figure}

}


\subsection{Computation Graphs of STL Operators}\label{subsec:graph_operation}
First, we consider the computation graph of each STL robustness formula given in Section \ref{subsec:prelim stl syntax and semantics} individually.

\subsubsection{Non-temporal STL Operators}
\label{subsubsec:nontemporal stl operators}
\begin{figure*}[t]
    \centering
    \subfloat[$\mathcal{G}_{\mu_c}$]{
    \label{fig:predicate graph}
    \includegraphics[width=35mm]{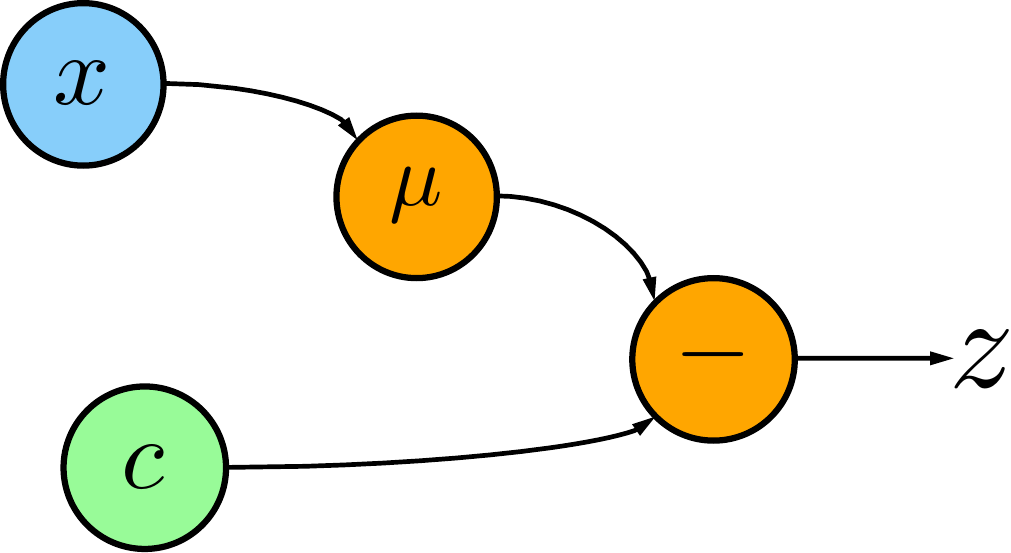}} $\quad$\subfloat[$\mathcal{G}_{\neg}$]{
    \label{fig:negation graph}
    \includegraphics[width=25mm]{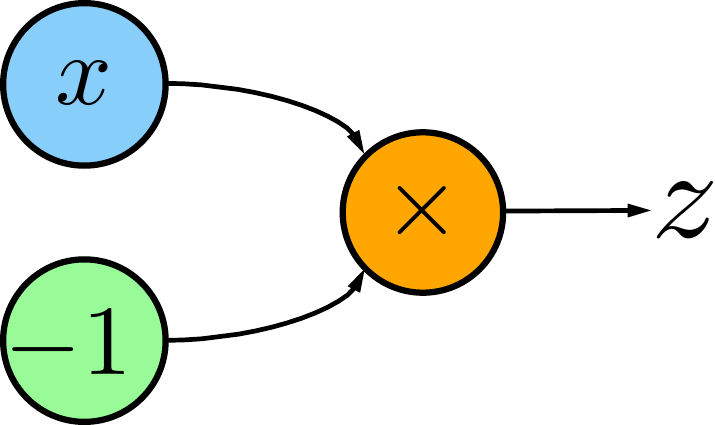}}$\quad$\subfloat[$\mathcal{G}_{\Rightarrow}$]{
    \label{fig:implies graph}
    \includegraphics[width=40mm]{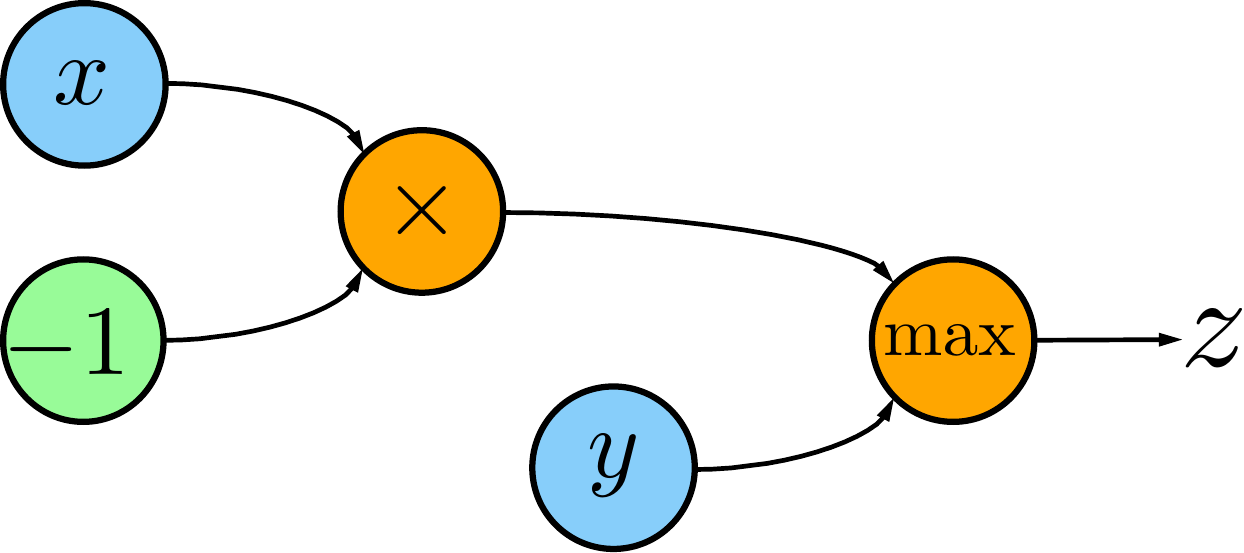}}$\quad$\subfloat[$\mathcal{G}_{\wedge}$]{
    \label{fig:and graph}
    \includegraphics[width=25mm]{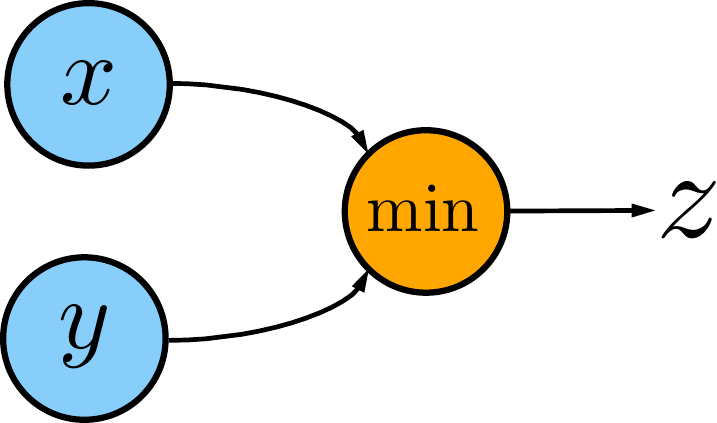}}$\quad$\subfloat[$\mathcal{G}_{\vee}$]{
    \label{fig:or graph}
    \includegraphics[width=25mm]{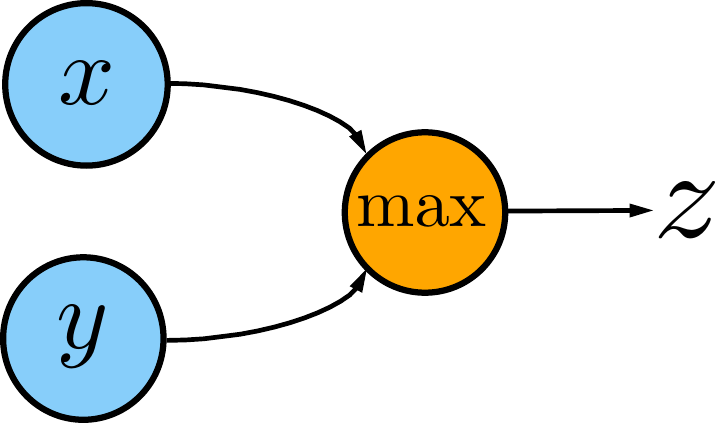}}
\caption{An illustration the computation graph representing the STL robustness formula for (a) predicates, (b) negation, (c) implies, (d) and, and (e) or. Blue nodes denote input variables, green nodes denote parameter values, and orange nodes denote mathematical operations.}
\label{fig:computation graph nontemporal}
\end{figure*}

The robustness formulas for the non-temporal operators are relatively straightforward to implement as computation graphs because they involve simple mathematical operations, such as subtraction, $\max$, and $\min$, and do not involve looping over time.
The computation graph for a predicate ($\mu_c$), negation ($\neg$), implies ($\Rightarrow$), and ($\wedge$), and or ($\vee$) are illustrated in Figure~\ref{fig:computation graph nontemporal}.
The input $x$ (and $y$) is either the signal of interest (for predicates) or the robustness trace(s) of a subformula(s) (for non-predicate operators).
While the output $z$ is the robustness trace after applying the corresponding STL robustness formula.
For example, when computing the robustness trace of $\phi \wedge \psi$ for a signal $s$, the inputs for $\mathcal{G}_{\wedge}$ are $\tau(s, \phi)$ and $\tau(s, \psi)$, and the output is $\tau(s, \phi \wedge \psi)$.

\subsubsection{Temporal STL Operators}
Due to the temporal nature of the eventually $\lozenge_{[a,b]}$, always $\square_{[a,b]}$, and until $\mathcal{U}_{[a,b]}$ operators, the computation graph construction needs to be treated differently than the non-temporal operators.
While computing the robustness \emph{value} of any temporal operators given an input signal is straightforward, computing the robustness \emph{trace} in a computationally efficient way is less so.
To accomplish this, we apply dynamic programming \citep{FainekosSankaranarayananEtAl2012} by using a recurrent computation graph structure, similar to the recurrent structure of recurrent neural networks (RNN) \citep{HochreiterSchmidhuber1997}. In doing so, we can compute the robustness value for every subsignal of an input signal.
The complexity of this approach is linear in the length of the signal for $\lozenge$ and $\square$, and at most quadratic for the $\mathcal{U}$ operator.

\subsubsection{Eventually and Always Operators}
\label{subsubsec:eventually and always}
\begin{figure}[t]
    \centering
    \includegraphics[width=0.4\textwidth]{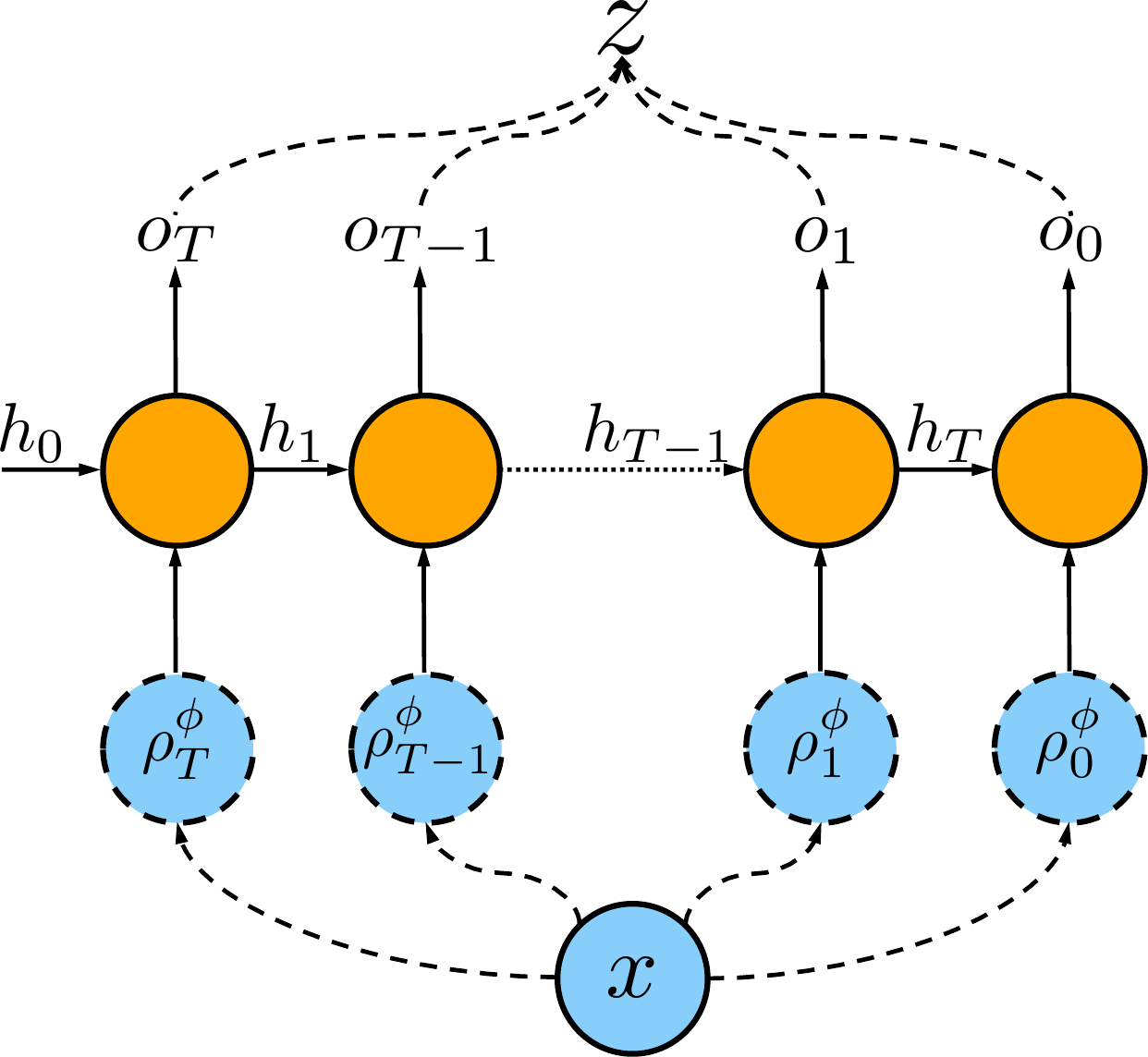}
    \caption{A schematic of an unrolled recurrent computation graph for the $\lozenge$ (eventually) and $\square$ (always) temporal operators, differing in the implementation of the recurrent cells depicted as orange circles. The dashed lines indicate the decomposition/reconstruction of the input/output signal via the time dimension. }
    \label{fig:temporal_cg}
\end{figure}

We first consider the Eventually operator, noting that similar a construction applies for the Always operator.
Let $\psi = \lozenge_{[a,b]} \phi$ \revised{and $s$ be a signal}; the goal is to construct the computation graph $\mathcal{G}_{\lozenge_{[a,b]}}$ which takes as input $\tau(s,\phi)$ (the robustness trace of the subformula) and outputs $\tau(s,\lozenge_{[a,b]}\phi)$.
For ease of notation, we denote $\rho(s_{t_i},\phi) = \rho_i^\phi$ as the robustness value of $\phi$ for subsignal $s_{t_i}$.
Figure~\ref{fig:temporal_cg} illustrates the unrolled graphical structure of $\mathcal{G}_{\lozenge_{[a,b]}}$.
\revised{Borrowing terminology from the RNN literature,} 
the $i$-th recurrent (orange) node takes in a hidden state $h_i$ and an input state $\rho_{T-i}^\phi$, and produces an output state $o_{T-i}$ (note the time indices are reversed in order to perform dynamic programming).
\revised{In the RNN literature, a ``hidden'' state is designed to summarize past input states and neural networks are used to update the hidden state given the previous hidden state and current input state. In this work, instead of using neural networks to update the hidden state, we use $\max$ or $\min$ depending on the STL operator of interest.}
By collecting all the $o_i$'s, the outputs of the computation graph form the (backward) robustness trace of $\psi$, i.e., $\tau(s,\psi)$.
The output robustness trace is treated as the input to another computation graph representing the next STL operator dictated by $\mathcal{G}$.


\begin{table*}[t]
    \centering
        \begin{tabular}{|c|cc|cc|cc|cc|cc|cc|}
        \hline
         & $h_0$ & $o_5$ & $h_1$ & $o_4$ & $h_2$ & $o_3$ & $h_3$ & $o_2$ & $h_4$ & $o_1$ & $h_5$ & $o_0$ \\ \hline
        $\lozenge_{[0, t_T]} (s > 0)$ & 1 & 1 & 1 & 3 & 3 & 3 & 3 & 3 & 3 & 3 & 3 & 3 \\
        $\lozenge_{[0, 2]} (s > 0)$ & (1, 1) & 1 & (1, 1) & 3 & (1, 3) & 3 & (3, 2) & 3 & (2, 1) & 2 & (1, 1) & 1 \\
        $\lozenge_{[2, t_T]} (s > 0)$ & (1,(1,1)) & 1 & (1,(1,1)) & 1 & (1,(1,3)) & 1 & (1,(3,2)) & 3 & (3,(2,1)) & 3 & (3,(1,1)) & 3 \\
        $\lozenge_{[1,3]} (s > 0)$ & (1, 1, 1) & 1 & (1, 1, 1) & 1 & (1, 1, 3) & 3 & (1, 3, 2) & 3 & (3, 2, 1) & 3 & (2, 1, 1) & 2 \\\hline
        \end{tabular}
    \caption{Hidden and output states of $\lozenge_\mathrm{I} (s > 0)$ for a signal $s = 1, 1, 1, 2, 3, 1$ and different time intervals $\mathrm{I}$. The signal is fed into the computation graph \emph{backward}.}
    \label{tab:stlcg example}
\end{table*}

Before describing the mechanics of the computation graph $\mathcal{G}_{\lozenge_{[a,b]}}$, we first define $M_N\in \mathbb{R}^{N\times N}$ to be a square matrix with ones in the upper off-diagonal and $B_N\in \mathbb{R}^N$ to be a vector of zeros with a one in the last entry. If $x\in\mathbb{R}^N$ and $u\in\mathbb{R}$, then the operation $M_Nx + B_Nu$ removes the first element of $x$, shifts all the entries up one index and replaces the last entry with $u$.
We distinguish four variants of $\mathcal{G}_{\lozenge_{[a,b]}}$ which depends on the interval $[a,b]$ attached to the temporal operator. \revised{Let $t_T$ denote the final time of a signal.}

\revised{\noindent\textbf{Case 1: $\lozenge\phi$}}. Let the initial hidden state be $h_0 = \rho_T^\phi$. The hidden state is designed to keep track of the largest value the graph has processed so far.
The output state for the first step is $o_T = \max(h_0,\rho_T^\phi)$, and the next hidden state is defined to be $h_1 = o_T$.
Hence the general update step becomes,
\[h_{i+1} = o_{T-i}, \quad o_{T-i} = \max(h_i, \rho_{T-i}^\phi).\]
By construction, the last step in this dynamic programming procedure,
\begin{align*}
    o_0 &= \max(h_T, \rho_0^\phi)\\
    o_0 &= \max(\max(h_{T-1}, \rho_1^\phi), \rho_0^\phi)\\
    & \vdots\\
    o_0 &= \max(h_0, \rho_T^\phi, \rho_{T-1}^\phi,\ldots,\rho_0^\phi)\\
    o_0 &= \rho(s_{t_0}, \lozenge\, \phi).
\end{align*}
The output states of $\mathcal{G}_{\lozenge \phi}$, $o_0, \, o_1, \ldots,\, o_T$, form $\tau(s, \lozenge \phi)$, the robustness trace of $\psi=\lozenge \phi$, and the last output state $o_0$ is $\rho(s,\lozenge\,\phi)$.

\revised{\noindent\textbf{Case 2: $\lozenge_{[0,b]}\,\phi$, $b < t_T$}}.  Let the initial hidden state be $h_0 = (h_{0_1},\, h_{0_2},\, \ldots ,\, h_{0_{N-1}})$, where $h_{0_i} = \rho_T^\phi,\:\forall i=1,...,N-1$\footnote{This corresponds to padding the input trace with the last value of the robustness trace. A different value could be chosen instead. This applies to Cases 3--4 as well.} and $N$ is the number of time steps contained in the interval $[0,b]$.
The hidden state is designed to keep track of inputs in the interval $(t_i, t_i + b]$.
The output state for the first step is $o_T = \max(h_0, \rho_T^\phi)$ (the $\max$ operation is over the elements in $h_0$ and $\rho_T^\phi$), and the next hidden state is defined to be $h_1 = M_{N-1}h_0 + B_{N-1}\rho_T^\phi$.
Hence the general update step becomes,
\begin{align*}
h_{i+1}  &= M_{N-1}h_i + B_{N-1}\rho_{T-i}^\phi,\\
 o_{T-i} &= \max(h_{i}, \rho_{T-i}^\phi).
\end{align*}
By construction, $o_0$ corresponds to the definition of $\rho(s_{t_0}, \lozenge_{[0,b]} \phi)$, $b < t_T$.

\revised{\noindent\textbf{Case 3: $\lozenge_{[a,t_T]}\,\phi$, $a > 0$}}. Let the hidden state be a tuple $h_i = (c_i, \,d_i)$. Intuitively, $c_i$ keeps track of the maximum value over future time steps starting from time step $t_i$, and $d_i$ keeps track of all the values in the time interval $[t_i,t_i + a]$. Specifically, $d_{i_1} = \rho_{t_i + a}^\phi$. Let the initial hidden state be $h_0 = (\rho_T^\phi, d_0)$ where $d_0 = (d_{0_1},\, d_{0_2},\, \ldots ,\, d_{0_{N-1}})$, $d_{0_i} = \rho_T^\phi, \: \forall i=1,...,N-1$ and $N$ is the number of time steps encompassed in the interval $[0,a]$. The output for the first step is $o_T = \max(c_0,\, d_{0_1})$, and the next hidden state is defined to be $h_1 = (o_T, \,M_{N-1}d_0 + B_{N-1}\rho_T^\phi$).
Following this pattern, the general update step becomes,
\begin{align*}
    h_{i+1} &= (o_{T-i}, \,M_{N-1}d_i + B_{N-1}\rho_{T-i}^\phi), \\
    o_{T-i} &= \max(c_i,\, d_{i_1}).
\end{align*}
By construction, $o_0$ corresponds to the definition of $\rho(s_{t_0}, \lozenge_{[a,t_T]} \phi)$, $a > 0$.

\revised{\noindent\textbf{Case 4: $\lozenge_{[a,b]}\,\phi$, $a > 0, \; b < t_T$}}. Let the initial hidden state be $h_0 = (h_{0_1},\, h_{0_2},\, \ldots ,\, h_{0_{N-1}})$, where $h_{0_i} = \rho_T^\phi, \: \forall i=1,...,N-1$  and $N$ is the number of time steps encompassed in the interval $[0,b]$.
The hidden state is designed to keep track of all the values between  $(t_i, t_i + b]$.
Let $M$ be the number of time steps encompassed in the $[a,b]$ interval.
The output for the first step is $o_T = \max(h_{0_1},\, h_{0_2},\, \ldots ,\, h_{0_{M}})$. The next hidden state is defined to be $h_1 = M_{N-1}h_0 + B_{N-1}\rho_T^\phi$.
Hence the general update step becomes,
\begin{align*}
h_{i+1} &= M_{N-1}h_i + B_{N-1} \rho_{T-i}^\phi,\\
o_{T-i} &= \max(h_{i_1},\, h_{i_2},\, \ldots ,\, h_{i_{M}})
\end{align*}
By construction, $o_0$ corresponds to the definition of $\rho(s_{t_0}, \lozenge_{[a,b]} \phi)$, $a > 0$, $b<t_T$.

The computation graph for the Always operation is the same but instead uses the $\min$ operation instead of $\max$.
The time complexity for computing $\tau(s, \lozenge_{[a,b]}\phi)$ (and also $\tau(s, \square_{[a,b]}\phi)$) is $\mathcal{O}(T)$ since it requires only one pass through the signal.
See Example~\ref{eg:stlcg example} for a concrete example.

\begin{example}\label{eg:stlcg example}
Let the values of a signal be $s = 1, 1, 1, 2, 3, 1$. Then the hidden and output states for $\lozenge_I$ with different intervals $I$ are given in Table~\ref{tab:stlcg example}.
\end{example}

\subsubsection{Until Operator}

Recall that the robustness formula for the Until operator is,
{\small
\[\rho(s_t, \phi \,\mathcal{U}_{[a,b]}\psi)  = \max_{t^\prime \in [t+a, t+b]} \min\lbrace\rho(s_{t^\prime}, \psi), \min_{\tau\in [t, t^\prime]} \rho(s_{\tau}, \phi)\rbrace.\]
}Using the definition of the Always robustness formula, the Until robustness formula can be rewritten as,
\begin{equation}
\rho(s_t, \phi \,\mathcal{U}_{[a,b]}\psi)  = \max_{t^\prime \in [t+a, t+b]} \min\lbrace\rho(s_{t^\prime}, \psi), \rho(s_{t}^{t^\prime}, \square \phi)\rbrace
\label{eq:until robustness formula with always}
\end{equation}
where \revised{(recall from Definition~\ref{def:signal})} $s_{t}^{t^\prime}$ denotes a signal consisting of values from $t$ to $t^\prime$.
Before we describe the computation needed for different instantiations of $[a,b]$, we first define the following notation.
Let $\overleftarrow{s}_{t}^{t_i}$ denote the \emph{reversed} signal of $s_{t}^{t_i}$.
We also want to note that the inputs into $\mathcal{G}_{\phi \mathcal{U}_{[a,b]}\psi}$ are $\tau(s_{t}, \phi)$ and $\tau(s_{t}, \psi)$. \revised{Let $t_T$ denote the final time of a signal.}

\revised{\noindent\textbf{Case 1: $\phi\,\mathcal{U}\,\psi$}}. 
Suppose we have a signal $s_{t_0}$. For each time step $t_i$, we can easily compute $\tau(s_{t_0}^{t_i}, \square \phi)$ (using the methodology outlined in Section~\ref{subsubsec:eventually and always}). Therefore, we can compute $\min\lbrace\rho(s_{t_i}, \psi), \rho(s_{t_j}^{t_i}, \square \phi)\rbrace$ for all subsignals of $s_{t_0}^{t_i}$. After looping through all $t_i$'s, we can perform the outer maximization in \eqref{eq:until robustness formula with always} and therefore produce $\tau(s_{t_0}, \phi \,\mathcal{U}\, \psi)$. Since we looped through all the values of $s_{t_0}$ once and computed the robustness trace of $\square\, \phi$ within each iteration, the time complexity is $\mathcal{O}(T^2)$ where $T$ is the length of the signal.

\revised{\noindent\textbf{Case 2: $\phi\,\mathcal{U}_{[a, b]}\,\psi$, $0 \leq a < b < t_T$.}}
First, loop through each time $t_i$. Each iteration of this loop will produce $\rho(s_{t_i}, \phi \,\mathcal{U}_{[a,b]}\psi)$, the robustness value for subsignal $s_{t_i}$.
For a given $t_i$, we can compute $\rho(s_{t_i}^{t^\prime}, \square \phi)$ for all $t^\prime \in [t_i + a, t_i + b]$ by computing the robustness trace of $\square\phi$ over $\overleftarrow{s}_{t_i}^{t_i+b}$ and then taking the the last $N$ values of robustness trace where $N$ is the number of time steps inside $[a, b]$. As such, we can compute $\rho(s_{t_i}, \phi \,\mathcal{U}_{[a,b]}\, \psi) = \max_{t^\prime \in [t_i+a, t_i+b]} \min\lbrace\rho(s_{t^\prime}, \psi), \rho(s_{t_i}^{t^\prime}, \square \phi)\rbrace$. Since we loop through all time steps in the signal, and compute the Always robustness trace over $\overleftarrow{s}_{t_i}^{t_i+b}$, the time complexity is $\mathcal{O}(TM)$ where $T$ is the length of the signal, and $M$ is the number of time steps contained in $[0,b]$.

\revised{\noindent\textbf{Case 3: $\phi\,\mathcal{U}_{[a, t_T]}\, \psi$, $0 \leq a < t_T$.}}
This is the same as Case 2 but instead we compute $\rho(s_{t_i}^{t^\prime}, \square \phi)$ for all $t^\prime \in [t_i + a, t_T]$ (practically speaking, we just consider up until the end of the signal). The time complexity in this case is $\mathcal{O}(T^2)$.

We have just described how we can compute the robustness trace of the Until operation by leveraging the computation graph construction of the Always operator and a combination of for loops, and $\max$ and $\min$ operations, all of which can be expressed using computation graphs.


\subsection{Calculating Robustness with Computation Graphs}
\label{subsec:graph_formula}

Since we can now construct the computation graph corresponding to each STL operator of a given STL formula, we can now construct the computation graph $\mathcal{G}$ by stacking together computation graphs according to $\mathcal{T}$, the corresponding parse tree.
The overall computation graph $\mathcal{G}$ takes a signal $s$ as its input, and outputs the robustness trace as its output.
The total time complexity of computing the robustness trace for a given STL formula is at most $\mathcal{O}(|\mathcal{T}|T^2)$ where $|\mathcal{T}|$ is the number of nodes in $\mathcal{T}$, and $T$ is the length of the input signal.

By construction, the robustness trace generated by the computation graph matches exactly the true robustness trace of the formula, and thus \texttt{stlcg} is correct by construction (see Theorem~\ref{thm:correctness of stlcg}).

\begin{theorem}[Correctness of \texttt{stlcg}]\label{thm:correctness of stlcg}
For any STL formula \(\phi\)
and any signal \(s\),
let \(\mathcal{G}_\phi\) be the computation graph produced by \texttt{stlcg}.
Then passing a signal $s$ through $\mathcal{G}_\phi$ produces the robustness trace $\tau(s, \phi)$.
\end{theorem}

\subsection{Smooth Approximations to \texttt{stlcg}}\label{subsec:graph_smooth}
Due to the recursion over $\max$ and $\min$ operations, there is the potential for many of the gradients be zero and therefore potentially leading to numerical difficulties.
To mitigate this, we can leverage smooth approximations to the $\max$ and $\min$ functions.
Let $x\in \mathbb{R}^n$ and $\beta\in\mathbb{R}_{\geq 0}$. We can use either the softmax/min or logsumexp approximations,

\begin{align*}
\widetilde{\max}_\beta(x) & = \frac{\sum_i^n x_i \exp(\beta x_i)}{\sum_j^n\exp(\beta x_j)} \: \text{or} \: \frac{1}{\beta}\log \sum_{i=1}^n \exp(\beta x_i)\\
\widetilde{\min}_\beta(x) & = \frac{\sum_i^n x_i \exp(-\beta x_i)}{\sum_j^n\exp(-\beta x_j)} \: \text{or} \: -\frac{1}{\beta}\log \sum_{i=1}^n \exp(-\beta x_i)
\end{align*}
where $x_i$ represents the $i$-th element of $x$, and $\beta $ operates as a scaling parameter. The approximation approaches the true solution when $\beta\rightarrow \infty$ while $\beta\approx0$ results in the mean value of the entries of $x$.
In practice, $\beta$ can be annealed over gradient-descent iterations.

Further, $\max$ and $\min$ are pointwise functions. As a result, the robustness of an STL formula can be highly sensitive to one single point in the signal, potentially providing an inadequate robustness metric especially if the signal is noisy \citep{MehdipourVasileEtAl2019}, or causing the gradients to accumulate to a single point only. We propose using an integral-based STL robustness formula $\mathcal{I}^M_{[a,b]}$ to ``spread out'' the gradients, and it is defined as follows.
For a given weighting function $M(t)$,
\[\rho(s_t,  \mathcal{I}^M_{[a,b]} \phi) = \sum_{\tau = t+a}^{t+b} M(\tau)\rho(s_\tau, \phi).\]
The integral robustness formula considers the weighted sum of the input signal over an interval $[a,b]$ and therefore produces a smoother robustness trace and the implication of this is demonstrated in Section~\ref{subsec:examples_motion_planning}.


\section{Case Studies: Using \texttt{stlcg} for Robotics Applications}\label{sec:examples}
We demonstrate the versatility and computational advantages of using \texttt{stlcg} by investigating a number of case studies in this section.
In these examples, we show (i) how our approach can be used to incorporate logical requirements into motion planning problems (Section \ref{subsec:examples_motion_planning}), (ii) the computational efficiency of \texttt{stlcg} achieved via parallelization and batching (Section \ref{subsec:examples_pstl}), and (iii) how we can use \texttt{stlcg} to translate human-domain knowledge into a form that can be integrated with deep neural networks (Section \ref{subsec:examples_neural_networks}). Code can be found at \url{https://github.com/StanfordASL/stlcg}.

\subsection{Motion Planning with STL Constraints}\label{subsec:examples_motion_planning}

Recently, there has been a lot of interest in motion planning with STL constraints (e.g., \cite{PantAbbasEtAl2017,RamanDonzeEtAl2014}); the problem of finding a sequence of states and controls that drives a robot from an initial state to a final state while obeying a set of constraints which includes STL specifications.
For example, while reaching a goal state, a robot may be required to enter a particular region for three time steps before moving to its final destination.
Rather than reformulating the problem as a MILP to account for STL constraints as done in \cite{RamanDonzeEtAl2014}, we consider a simpler approach of augmenting the loss function with an STL robustness term, and therefore enable a higher degree of customization when designing how the STL specifications are factored into the problem.
\revised{We present two motion planning examples that uses a variety of STL operators.}

\begin{figure*}[h]
    \centering
    \subfloat[Trajectories satisfying $\phi_1$ and $\phi_2$ which uses the Always operator.]{
    \label{fig:motion planning trajectories Always}
    \includegraphics[width=0.37\textwidth]{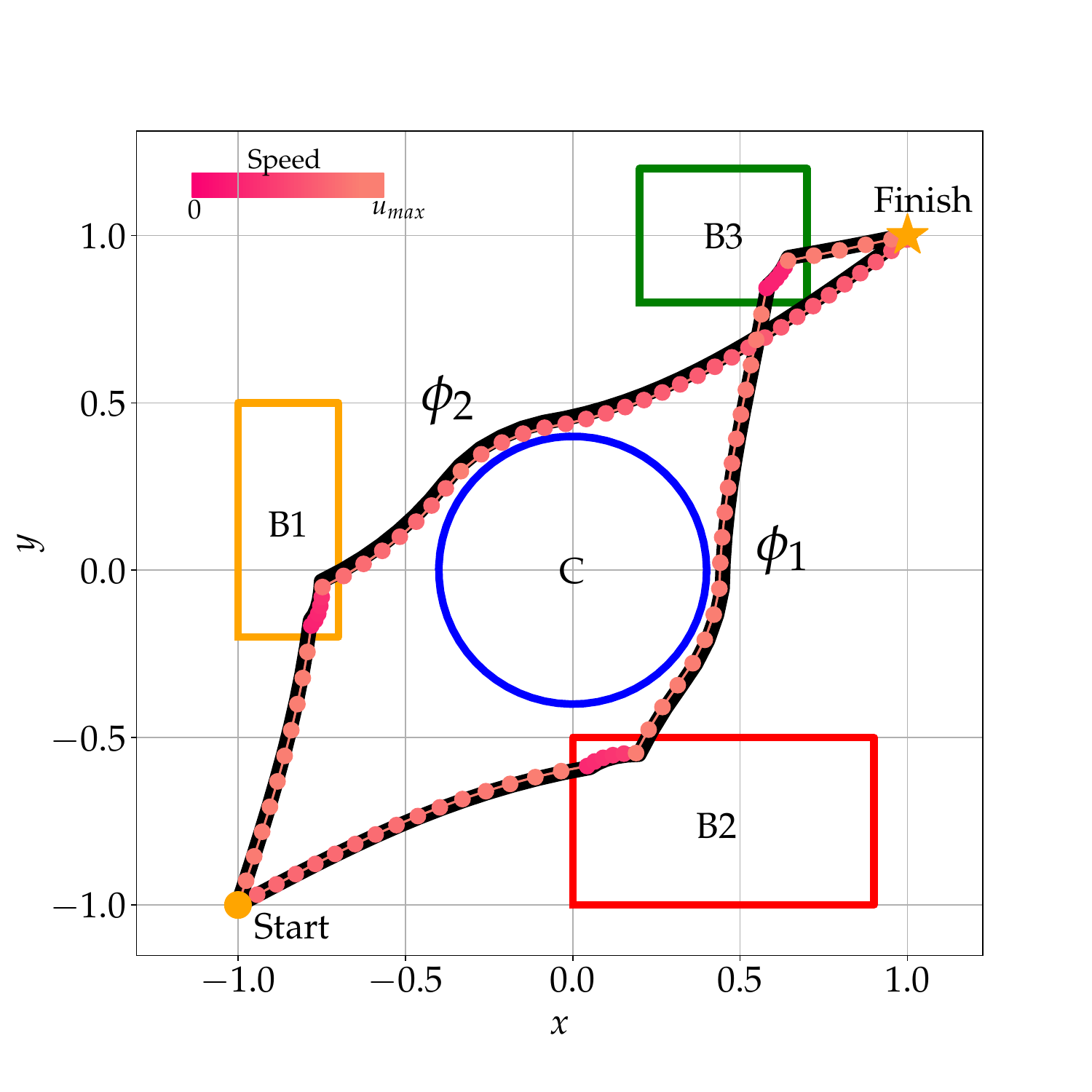}} $\qquad \qquad$ \subfloat[Trajectories satisfying $\psi_1$ and $\psi_2$ which uses the integral operator.]{
    \label{fig:motion planning trajectories Integral}
    \includegraphics[width=0.37\textwidth]{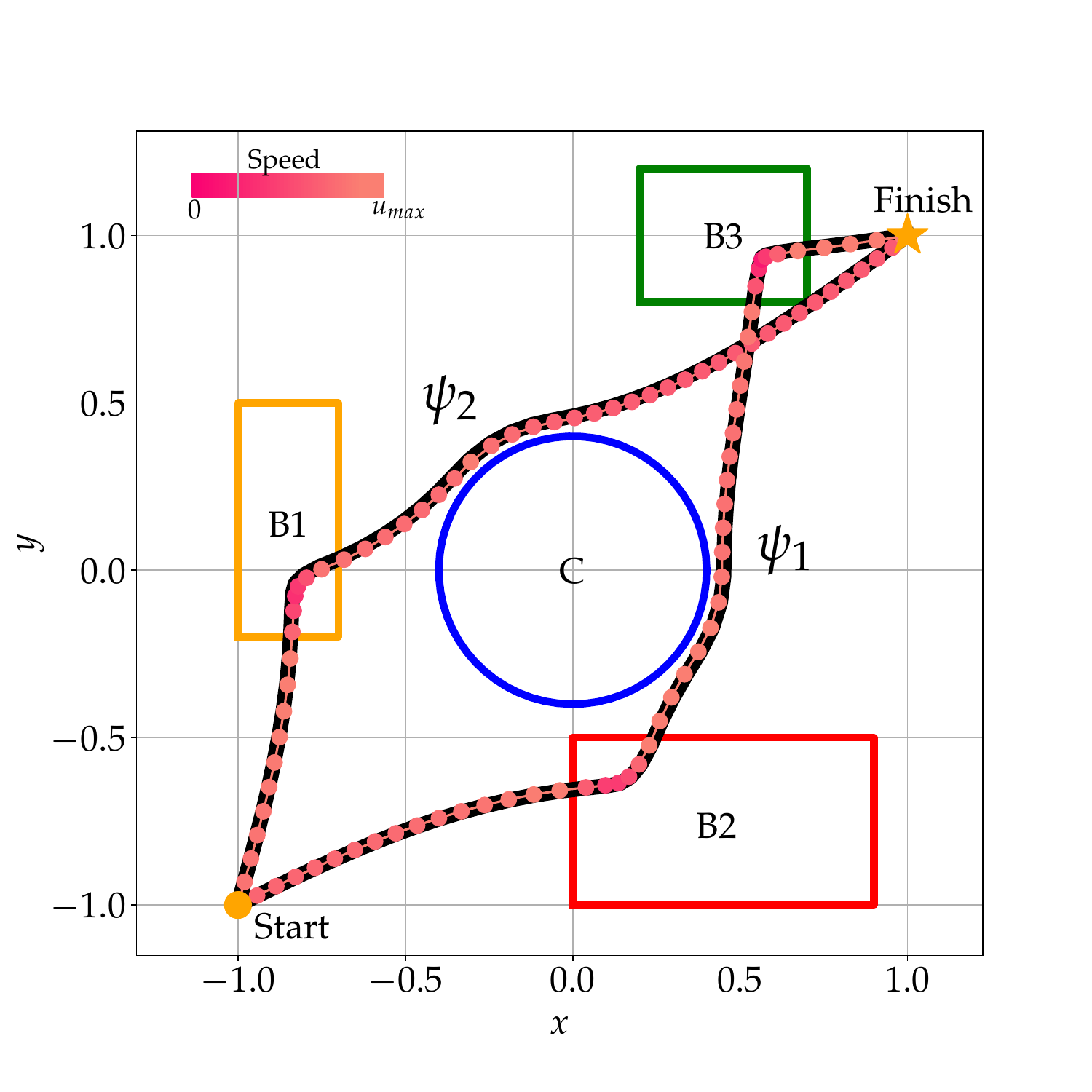}}
\caption{\revised{State trajectories from solving the motion planning problem described in Section~\ref{subsubsec:motion planning 1}.}}
\label{fig:motion planning example}
\end{figure*}

\begin{figure}[h]
    \centering
    \includegraphics[width=0.45\textwidth]{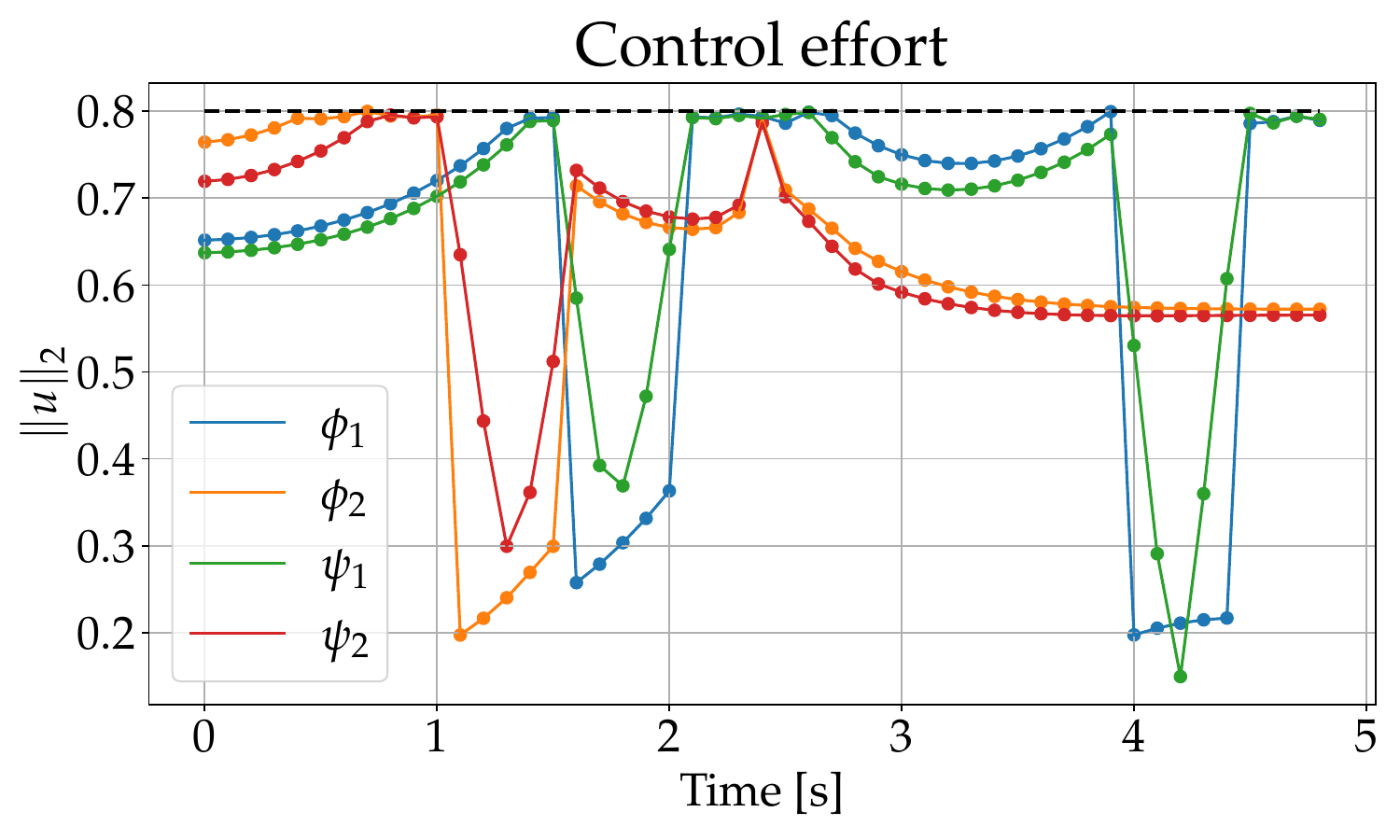}
    \caption{\revised{Control trajectories from solving the motion planning problem described in Section~\ref{subsubsec:motion planning 1}. All cases satisfy $\theta =\square \|u\|_2 \leq 0.8$.}}
    \label{fig:motion planning controls}
\end{figure}

\subsubsection{Motion Planning Example 1}\label{subsubsec:motion planning 1}

Consider the following motion planning problem illustrated in Figure~\ref{fig:motion planning example}; a robot must find a sequence of states $X = x_{1:N}$ and controls $U = u_{1:N-1}$ that takes it from the yellow circle (position = $(-1,-1)$) to the yellow star (position=$(1,1)$) while satisfying an STL constraint $\phi$.
Assume the robot is a point mass with state $x \in \mathbb{R}^2$ and action $u\in\mathbb{R}^2$. It obeys single 2D integrator dynamics $\dot{x} = u$ and has a control constraint $\|u\|_2 \leq u_{\max}$.
We can express the control constraint as an STL formula: $\theta = \square \, \|u\|_2 \leq  u_{\max}$.

This motion planning problem can be cast as an unconstrained optimization problem \revised{which minimizes the following cost function},
\begin{align*}
    \min_{X,\,U} \: \| Ez - D\|_2^2 + \gamma_1 J_m(\rho(X, \phi)) + \gamma_2 J_m(\rho(U, \theta))
\end{align*}
where $z = (X,U)$ is the concatenated vector of states and controls \revised{over all time steps}.
Since the dynamics are linear, we can express the dynamics, and start and end point constraints across all time steps in a single linear equation $Ez = D$.
The other function, $J_m$, represents the cost on the STL robustness value with margin $m$. We use $J_m(x) = \mathrm{ReLU}(-(x - m))$ where $\mathrm{ReLU}(x)=\max(0,x)$ is the rectified linear unit. This choice of $J_m$ incentivizes the robustness value to be greater or equal to the margin $m$.

To showcase a variety of STL constraints, we consider four different STL specifications,
\begin{align*}
\phi_1 &= \lozenge\square_{[0,5]} \text{inside B2}  \wedge  \lozenge  \square_{[0, 5]} \text{inside B3}  \wedge  \square \neg  \text{inside C}\\
\phi_2 &= \lozenge\square_{[0,5]} \text{inside B1}  \wedge  \square \neg \text{inside B3}  \wedge  \square \neg  \text{inside C}.\\
\psi_1 &= \lozenge\mathcal{I}^{\Delta t^{-1}}_{[0,5]} \text{inside B2}  \wedge  \lozenge  \mathcal{I}^{\Delta t ^{-1}}_{[0, 5]} \text{inside B3}  \wedge  \square \neg  \text{inside C}\\
\psi_2 &= \lozenge\mathcal{I}^{\Delta t ^{-1}}_{[0,5]} \text{inside B1}  \wedge  \neg \mathcal{I}^{\Delta t ^{-1}} \text{inside B3}  \wedge  \square  \neg \text{inside C},
\end{align*}
where B1, B2, B3, and C are the regions illustrated in Figure~\ref{fig:motion planning example}.
$\phi_1$ translates to: the robot needs to eventually stay inside the red box (B2) and green box (B3) for five time steps each, and never enter the blue circular region (C). Similarly, $\phi_2$ translates to: the robot eventually needs to stay inside the orange box (B1) for five time steps, and never enter the green box (B3) nor the blue circular region (C).
$\psi_1$ and $\psi_2$ are similar to $\phi_1$ and $\phi_2$ except that the integral robustness formula (described in Section \ref{subsec:graph_smooth}) is used.

We initialize the solution with a straight line connecting the start and end points, noting that this initialization violates the STL specification. We then perform standard gradient descent (constant step size of 0.05) with $\gamma_1 = \gamma_2 = 0.3$, $m=0.05$, and $\Delta t^{-1} = \frac{1}{0.1}$; the solutions are illustrated in Figure~\ref{fig:motion planning example}.
As anticipated, using the integral robustness formula results in a smoother trajectory and smoother control signal than using the Always robustness formula. We also note that the control constraint ($u_{\max} = 0.8$) was satisfied for all cases (see Figure~\ref{fig:motion planning controls}).

\revised{

\begin{figure*}[tb]
    \centering
    \includegraphics[width=\textwidth]{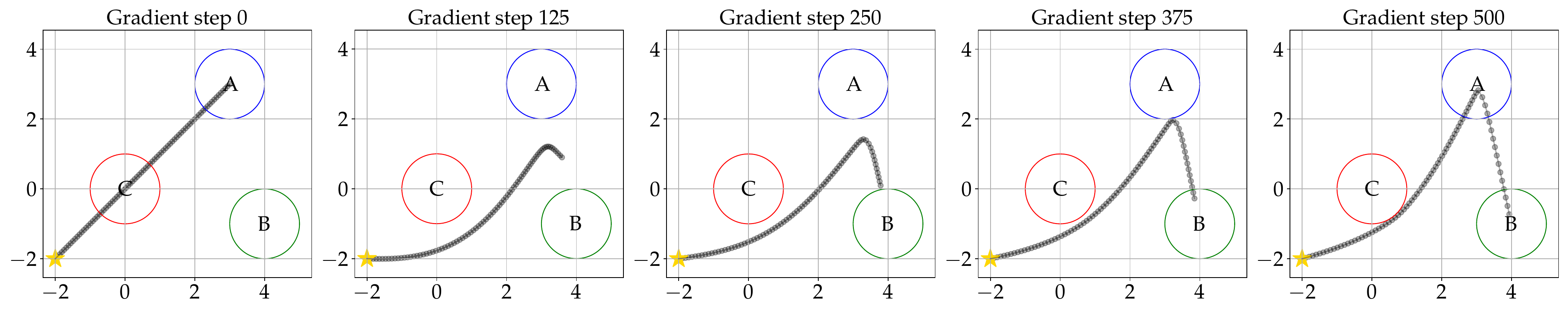}
    \caption{\revised{State trajectory at various gradient steps during the gradient descent process for solving the motion planning problem described in \eqref{eq:motion planning example 2}.}}
    \label{fig:motion_planning_until_example}
\end{figure*}

\begin{figure}[tb]
    \centering
    \includegraphics[width=0.45\textwidth]{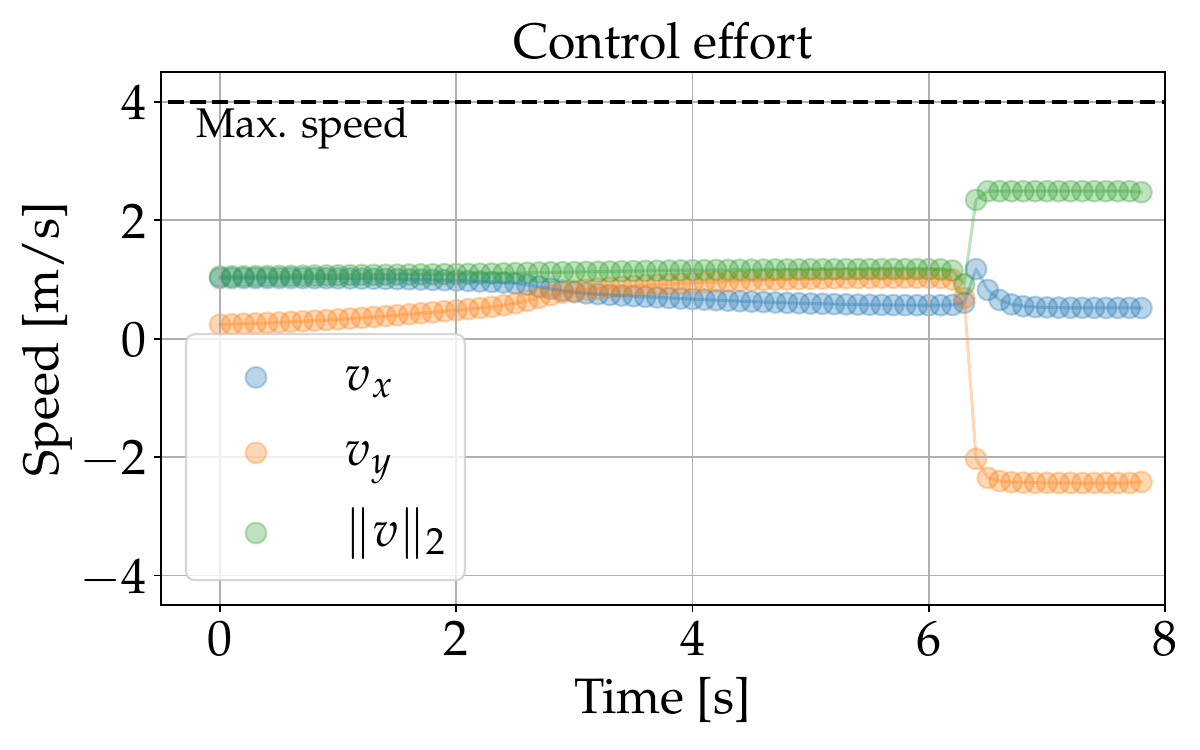}
    \caption{\revised{Control trajectory solution from performing gradient descent on a motion planning problem described in \eqref{eq:motion planning example 2}.}}
    \label{fig:motion_planning_until_example_control}
\end{figure}

}

\revised{
\subsubsection{Motion Planning Example 2}

In this example, we consider an STL specification containing the Until operator which specifies a temporal ordering in its STL subformulas. On the other hand, the STL formulas studied in the previous examples do not specify any temporal ordering, but rather the ordering in the final solution depends on the initial guess.
Recall that a formula with the Until operator, $\phi = \phi_1 \mathcal{U} \phi_2$, requires that $\phi_1$ be true before $\phi_2$ is true (see Section~\ref{subsec:prelim stl syntax and semantics}). For instance, an Until operator could specify a robot to first visit region A before region B even though visiting region B first may result in a shorter path, or is more similar to the trajectory used as the initial guess for the trajectory optimization problem.

Using the same system dynamics in the previous example (2D single integrator dynamics), we solve the following motion planning problem for the environment illustrated in Figure~\ref{fig:motion_planning_until_example}: \emph{Starting at $[-2, -2]^T$, the robot must first visit region A and then stay in region B while avoiding region C and stay below a fixed speed limit.}
For a fixed maximum number of time steps $T$ and time step size $\Delta t$, let $X=x_{0:T}$ and $U=u_{0:T-1}$ denote the state (i.e., position) and control (i.e., velocity) trajectory over all time steps. We can solve the above motion planning problem via the following unconstrained optimization problem:

\begin{align}
 \max_U \quad&\lambda_1\rho(X, \phi) + \lambda_2 \rho(U, \phi_\mathrm{u}) - \lambda_3\mathrm{ReLU}(-\rho(X, \phi_\mathrm{obs}))\notag\\
\mathrm{where}\: & X = x_{0:T},\, U = u_{0:T-1}\notag\\
& x_0 = [-2, -2]^T\notag \\
& x_{t+1} = x_t + u_t\Delta t,\quad t = 0,...,T-1\notag\\
& \phi = (\lozenge \,\text{inside region A} ) \,\mathcal{U} (\lozenge \square\, \text{inside region B})\notag\\
& \phi_\mathrm{u} = \square\,  \|u\|_2 < u_{\max}\notag\\
& \phi_\mathrm{obs} = \square \, (\neg \,\text{inside region C}).\label{eq:motion planning example 2}
\end{align}

Note that the last term in the objective function does not incentivize the solution to avoid the obstacle more than necessary, i.e., there is only a penalty for entering the obstacle, but no additional reward is given for being far away from it.

Figures~\ref{fig:motion_planning_until_example} and \ref{fig:motion_planning_until_example_control} show the results from solving \eqref{eq:motion planning example 2} via gradient descent with
$T = 79$, $\Delta t = 0.1$, $\lambda_1 = \lambda_3 =1$, $\lambda_2 = 0.5$, $u_{\max} = 4$ms$^{-1}$, and 500 gradient steps with step size of 0.5.
We see that the trajectory is initialized with a trajectory that ends inside region A and enters region C, therefore violating the STL specifications.
After performing multiple gradient descent steps, the trajectory is able to satisfy the Until, obstacle avoidance, and speed limit STL specifictions.

}

\begin{figure*}[t]
    \centering
    \includegraphics[width=0.9\textwidth]{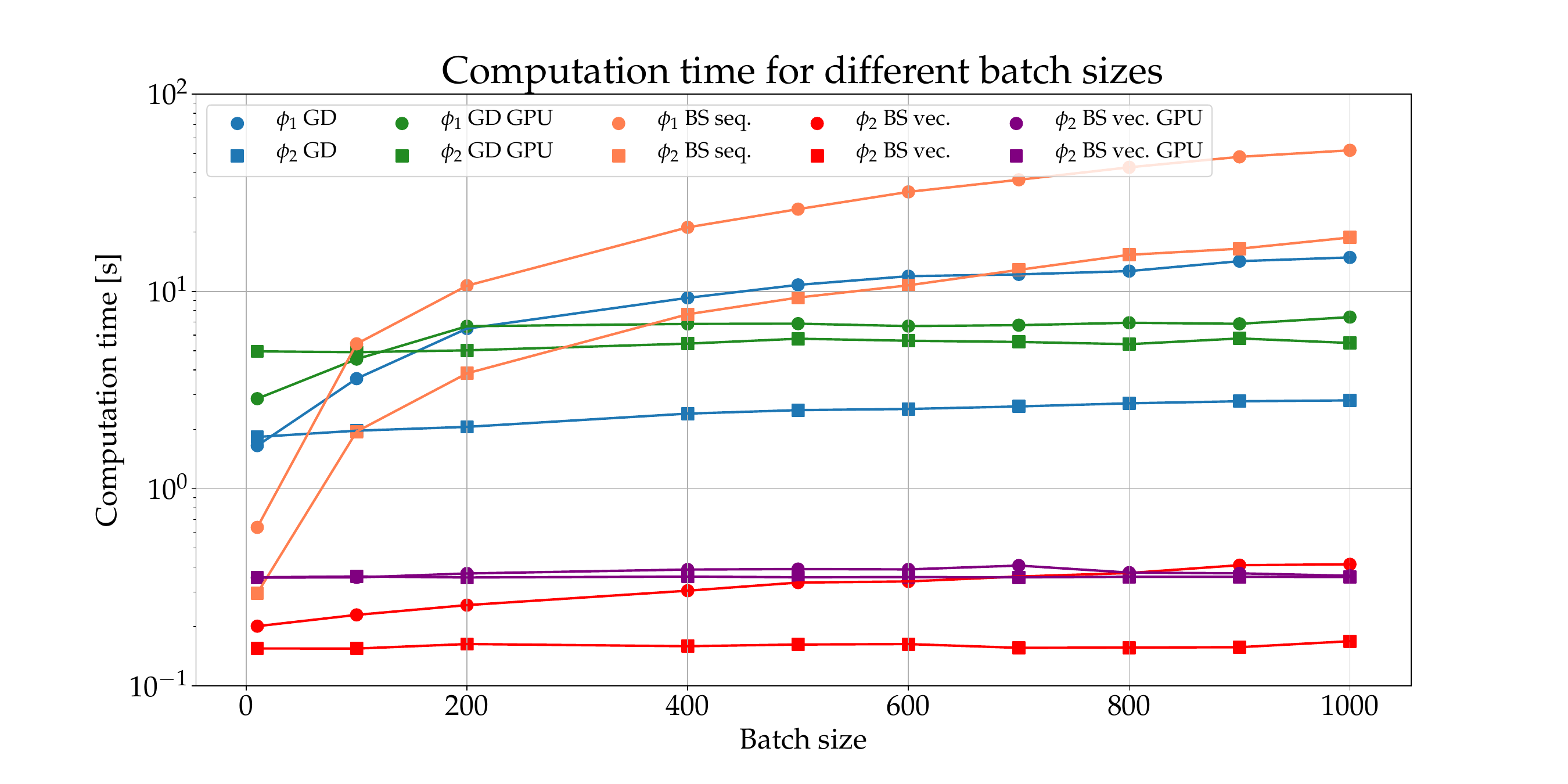}
    \caption{\revised{Computation time of using gradient descent (GD) and binary search (BS) with and without GPU utilization to solve pSTL problems. For the binary search approach, a sequential (seq.) and vectorized (vec.) approach were considered.} The experiments were  computed using a 3.0GHz octocore AMD Ryzen 1700 CPU and a Titan X (Pascal) GPU. }
    \label{fig:clustering computation time}
\end{figure*}

\begin{figure*}[t]
    \centering
    \includegraphics[width=0.95\textwidth]{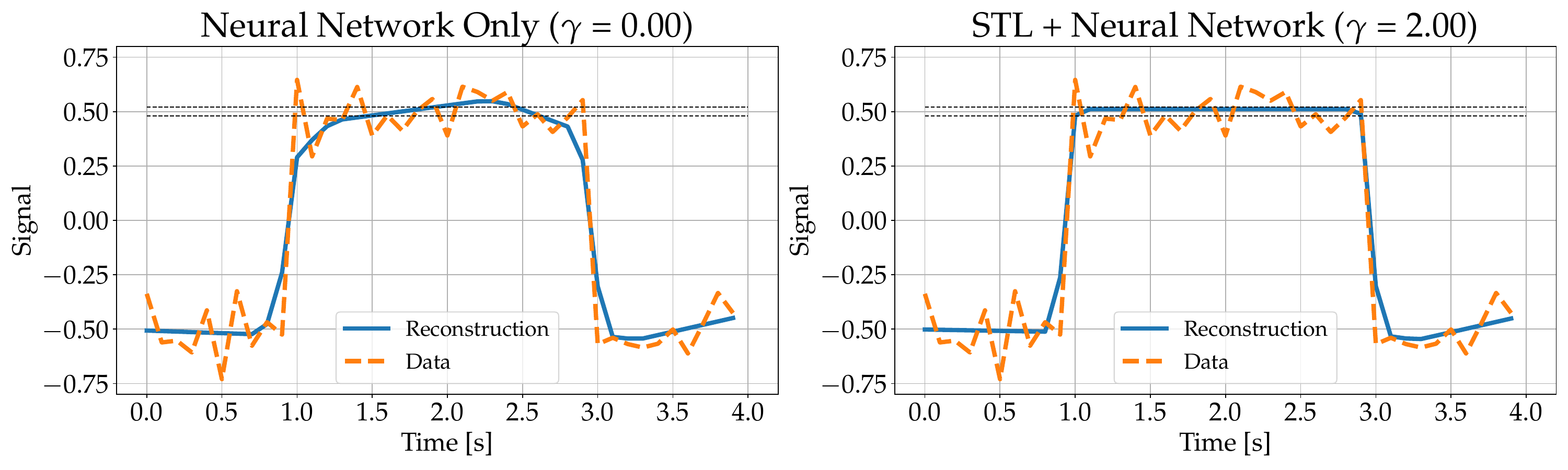}
    \caption{A simple supervised learning problem without (left) and with (right) STL regularization. The output is required to satisfy $\phi = \square_{[1,2.75]} (s > 0.48 \, \wedge \, s < 0.52)$ despite the training data being noisy and violating $\phi$.}
    \label{fig:bump example}
\end{figure*}

\subsection{Parametric STL for Behavioral Clustering}
\label{subsec:examples_pstl}

In this example, we consider parametric STL (pSTL) problems \citep{Vazquez-ChanlatteDeshmukhEtAl2017,AsarinDonzeEtAl2012},
and demonstrate that since modern automatic differentiation software such as PyTorch can be used to implement \texttt{stlcg}, we have the ability to very easily parallelize the computation and leverage GPU hardware.
The pSTL problem is a form of logic-based parameter estimation for time series data. It involves first constructing an STL template formula where the predicate parameter values and/or time intervals are unknown.
The goal is to find parameter values that best fit a given signal. \revised{For example, find parameters that result in zero robustness.}
As such, the pSTL process can be viewed as a form of feature extraction for time-series data. Logic-based clustering can be performed by clustering on the extracted pSTL parameter values. For example, pSTL has been used to cluster human driving behaviors in the context of autonomous driving applications \citep{Vazquez-ChanlatteDeshmukhEtAl2017}.

The experimental setup for our example is as follows. Given a dataset of $N$ step responses from randomized second-order systems, we use pSTL to help cluster different types of responses, e.g., under-damped, critically damped, or over-damped. Based on domain knowledge of second order step responses, we design the following pSTL formulas,
\begin{flalign*}
\qquad \phi_1 &=\square_{[50,100]}\, | s - 1| < \epsilon_1 & \text{\textit{(final value)}} \qquad\\
\qquad \phi_2 &= \square \, s < \epsilon_2  & \text{\textit{(peak value)}}\qquad
\end{flalign*}
For each signal $s^{(j)},\, j=1,\,\ldots,\,N$, we want to find $\epsilon_{1j}$ and $\epsilon_{2j}$ ($\epsilon_1$ and $\epsilon_2$ for the $j$th signal) that provides the best fit, i.e., robustness equals zero. Using \texttt{stlcg}, we can \emph{batch} \revised{or \emph{vectorize}} our computation and hence solve for all $\epsilon_{ij}$ more efficiently.\footnote{We can even account for variable signal length by padding the inputs and keeping track of the signal lengths.}
\revised{Additionally, we can very easily, with very minor code changes, make use of the GPU to potentially speed up the computation.

We experiment with two different approaches, (i) gradient descent, and (ii) binary search \citep{Vazquez-ChanlatteDeshmukhEtAl2017}. We note that we are simply running experiments using both approaches to illustrate the computational benefits of \texttt{stlcg}, and that we are not advocating for and against the method itself. Naturally, the preferred method of choice depends on the problem a user is interested in.}

\revised{
For the gradient descent method, we solve an optimization problem for each pSTL formula $\phi_i,\,i=1,2$. Since each signal is independent, we can batch the signal together and solve the pSTL problem for all signals simultaneously using a combined loss function. We use the following loss function,
$\min_{\epsilon_{ij}, j=1,..,N} \:\: \sum_{j=1}^N\, \mathrm{ReLU}(-\rho(s^{(j)}, \phi_i))$. Since the $\mathrm{ReLU}(-x)$ function results in zero gradient when the robustness is positive, gradient descent steps will only be performed on pSTL formulas until the robustness is non-negative. We perform experiments with and without using a GPU.

The binary search method described in \cite{Vazquez-ChanlatteDeshmukhEtAl2017} is applicable to monotonic pSTL formulas, formulas where the robustness value increases monotonically as the parameters increases (or decreases).
Since $\phi_1$ and $\phi_2$ are monotonic pSTL formulas, we investigate the average computation time taken to find a solution to all $\epsilon_{ij}$'s using the binary search method. 
In particular, we consider three cases, (i) \emph{sequential}: solves the binary search problem for each signal sequentially, (ii) \emph{vectorized}: vectorizes the signals and pSTL formulas (using \texttt{stlcg}) and then binary search is executed in a vectorized fashion, and (iii) \emph{vectorized + GPU}: the same as (ii) but uses GPU instead of the CPU.  
For all the experiments, the $\epsilon_{ij}$'s are initialized to zero, corresponding to parameters that result in negative robustness.
}

Both approaches converged to the same solution, and the resulting computation times are illustrated in Figure~\ref{fig:clustering computation time}.
\revised{Naturally, the computation time for executing binary search sequentially increases linearly and is the least scalable of the methods investigated.
By batching or vectorizing the inputs using \texttt{stlcg}, the computation time scales more efficiently. Further, GPU parallelization provides near-constant time computation and is more beneficial for larger problem sizes.}


For cases where the solution has multiple local minima (e.g., non-monotonic pSTL formulas), \revised{we can no longer use binary search. Instead, we could use gradient descent and} additionally batch the input with \revised{multiple} samples from the parameter space, and anneal $\beta$, the scaling parameter for the $\max$ and $\min$ approximation over each iteration. The samples will converge to a local minimum, and, with sufficiently many samples and an adequate annealing schedule, we can (hopefully) find the global minimum.
However, we note that we are currently not able to optimize time parameters that define an interval as we cannot backpropagate through those parameters. Future work will investigate how to address this, potentially leveraging ideas from forget gates in long short-term memory (LSTM) networks \cite{GersSchmidhuberEtAl1999}.

\subsection{Robustness-Aware Neural Networks}
\label{subsec:examples_neural_networks}
We demonstrate via a number of small yet illustrative examples how \texttt{stlcg} can be used to make learning-based models (e.g., deep neural networks) more robust and reflective of desired behaviors stemming from human-domain knowledge.

\begin{figure}[t]
    \centering
    \includegraphics[width=0.45\textwidth]{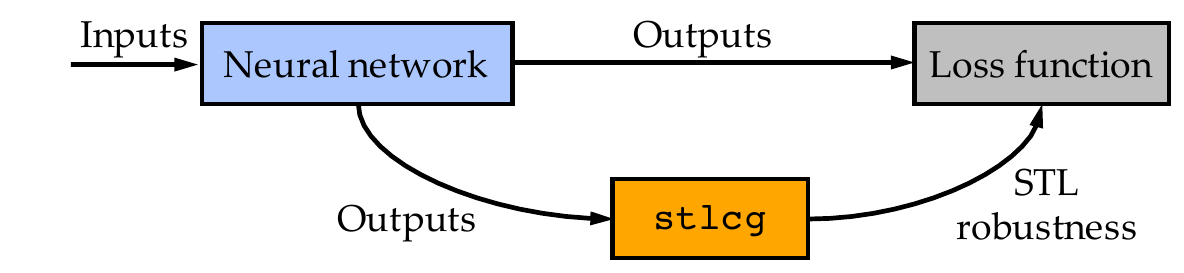}
    \caption{Architecture of a neural network enhanced with STL regularization in its loss function.}
    \label{fig:neural network arch}
\end{figure}

\subsubsection{Model Fitting with Known Structure}
In this example, we show how \texttt{stlcg} can be used to regularize a neural network model to prevent overfitting to the noise in the data.
Consider the problem of using a neural network to model temporal data such as the (noisy) signal shown by the orange line in Figure~\ref{fig:bump example}. Suppose that based on domain knowledge of the problem, we know that the data must satisfy the STL formula $\phi = \square_{[1,2.75]} (s > 0.48 \, \wedge \, s < 0.52)$.
Due to the noise in the data, $\phi$ is violated.
Neural networks are prone to overfitting and may unintentionally learn this noise.
We can train a single layer neural network $f_\theta$ to capture the bump-like behavior in the noisy data. However, this learned model violates $\phi$ (see Figure~\ref{fig:bump example} (left)).
To mitigate this, we regularize the learning processing by augmenting the loss function with a term that penalizes negative robustness values. Specifically, the loss function becomes
\begin{align}
\mathcal{L} = \mathcal{L}_0 + \gamma \mathcal{L}_\mathrm{STL}
\label{eq:loss function with robustness regularization}
\end{align}
where $\mathcal{L}_0$ is the original loss function (e.g., reconstruction loss), and, in this example, $\mathcal{L}_\mathrm{STL} = \mathrm{ReLU}(-\rho(f_\theta(x), \phi))$ is the robustness loss evaluated over the neural network output, and $\gamma = 2$. A schematic representing how, in general, a neural network can be regularized using \texttt{stlcg} is illustrated in Figure~\ref{fig:neural network arch}.

As shown in Figure~\ref{fig:bump example} (right), the model with robustness regularization is able to obey $\phi$ more robustly. Note that regularizing the loss function with an STL loss does not guarantee that the output will satisfy $\phi$ but rather the resulting model is encouraged to violate $\phi$ as little as possible, and the incentive can be controlled by the parameter $\gamma$.

\begin{figure*}[t]
    \centering
    \includegraphics[width=0.95\textwidth]{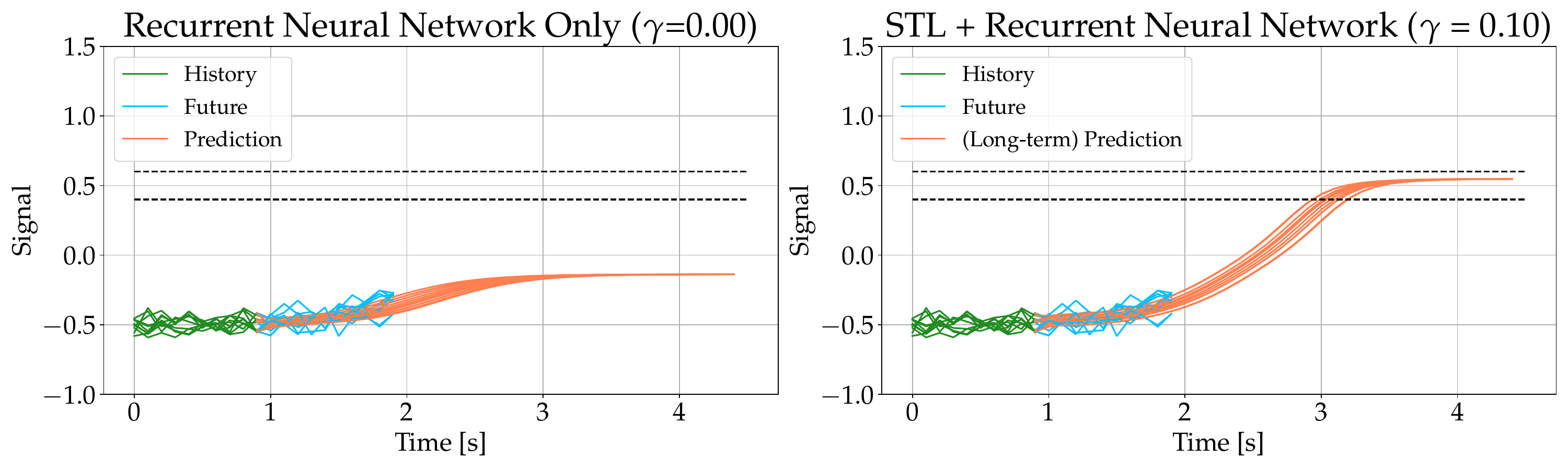}
    \caption{Comparison of a sequence-to-sequence prediction model without (left) and with (right) robustness regularization. With robustness regularization, the model achieves desired long-term behavior $\phi = \lozenge\, \square_{[0,1]} \, (s > 0.4 \, \wedge \, s < 0.6)$ despite having access to only short-term behaviors during training.}
    \label{fig:prediction example}
\end{figure*}

\subsubsection{Sequence-to-Sequence Prediction}
In this example, we demonstrate how \texttt{stlcg} can be used to influence long-term behaviors of sequence-to-sequence prediction problems to reflect domain knowledge despite only having access to short-term data.
Sequence-to-sequence prediction models are often used in robot decision-making, such as in the context of model-based control where a robot may predict future trajectories of other agents in the environment given past trajectories, and use these predictions for decision-making and control \citep{SchmerlingLeungEtAl2018}.
Often contextual knowledge of the environment is not explicitly labeled in the data, for instance, cars always drive on the right side of the road, or drivers tend to keep a minimum distance from other cars. \texttt{stlcg} provides a natural way, in terms of language and computation, to incorporate contextual knowledge into the neural network training process, thereby infusing desirable behaviors into the model which, as demonstrated through this example, can improve long-term prediction performance.

\begin{figure}[t]
    \centering
    \includegraphics[width=0.5\textwidth]{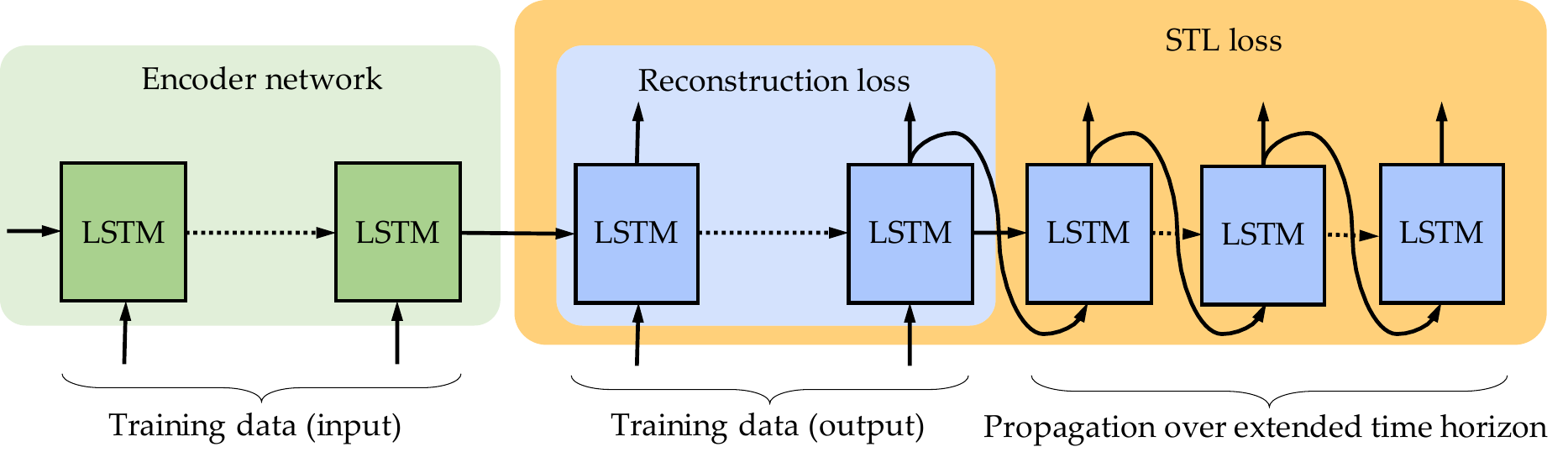}
    \caption{Schematic of a rolled-out RNN network illustrating how to leverage \texttt{stlcg} to improve long-term predictions with access to only short-term data.}
    \label{fig:rnn arch}
\end{figure}

We generate a dataset of signals $s$ by adding noise to the $\tanh$ function with random scaling and offset. Illustrated in Figure~\ref{fig:rnn arch}, we design an LSTM encoder that takes the first 10 time steps as inputs (i.e., trajectory history). The encoder is connected with an LSTM decoder network that predicts the next 10 time steps (i.e., future trajectory). Note, the time steps are of size $\Delta t = 0.1$.
Suppose, based on contextual knowledge, we know \emph{a priori} that the signal will eventually, beyond the next ten time steps captured in the data, be in the interval $[0.4, 0.6]$. Specifically, the signal will satisfy  $\phi = \lozenge\, \square_{[0,1]} \, (s > 0.4 \, \wedge \, s < 0.6)$.
We can leverage knowledge of $\phi$ by rolling out the RNN model over an extended time horizon beyond what is provided in the training data, and apply the robustness loss on the rolled-out signal $s^{(i)}$ corresponding to the input $x^{(i)}$.
We use \eqref{eq:loss function with robustness regularization} as the loss function where $\mathcal{L}_0$ is the mean square error reconstruction loss over the first ten predicted time steps, and $\mathcal{L}_\mathrm{STL}=\sum_i\mathrm{ReLU}(-\rho(s^{(i)}, \phi))$ is the total amount of violations over the extended roll-out.
With $\gamma = 0.1$, Figure~\ref{fig:prediction example} illustrates that with STL robustness regularization, the RNN model can successfully predict the next ten time steps and also achieve the desired long-term behavior despite being only trained on short-horizon data.

\subsubsection{Latent Space Structure}
\begin{figure}[t]
    \centering
    \includegraphics[width=0.45\textwidth]{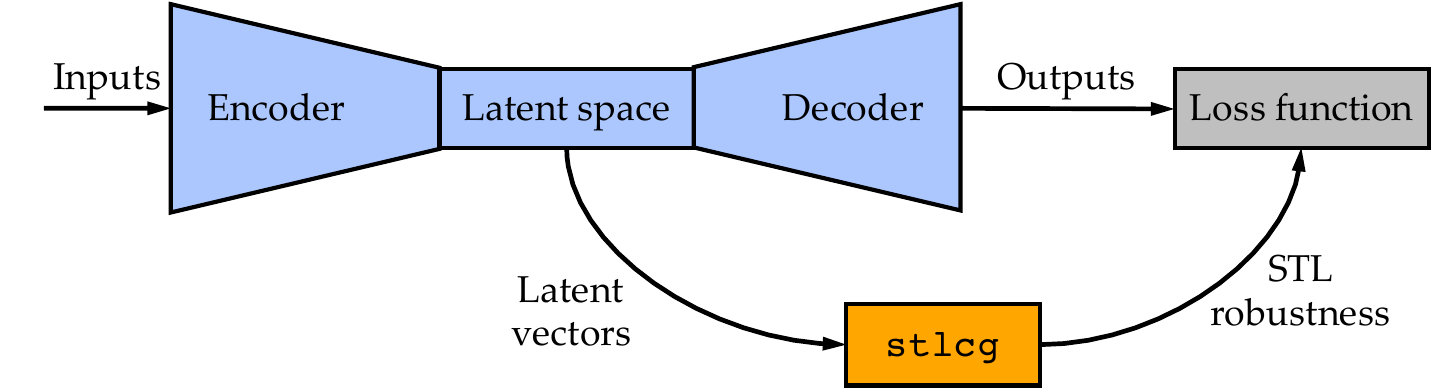}
    \caption{Architecture of a latent space model leveraging \texttt{stlcg} for added structure.}
    \label{fig:latent space arch}
\end{figure}

In this example, we demonstrate how \texttt{stlcg} can be used to enforce logic-based interpretable structure into deep latent space models and thereby improve model performance.
Deep latent space models are a type of neural network models that introduce a bottleneck in the architecture to force the model to learn salient (i.e., latent) features from the input in order to (potentially) improve modeling performance and provide a degree of interpretability. Specifically, an encoder network transforms the inputs into latent variables, which have a lower dimension than the inputs, and then a decoder network transforms the latent variables into desired outputs. An illustration of an encoder-decoder architecture is shown in Figure~\ref{fig:latent space arch} in blue.

Variational autoencoders (VAEs) \citep{KingmaWelling2013} are a type of probabilistic deep latent space models used in many domains such as image generation, and trajectory prediction.
Given a dataset, a VAE strives to learn a distribution such that samples drawn from the distribution will be similar to what is represented in the dataset (e.g., given a dataset of human faces, a VAE can generate new similarly-looking images of human faces).
For brevity, we omit describing details of VAEs but refer the reader to \cite{Doersch2016,JangGuEtAl2017} for an overview on VAEs.

Unfortunately, the VAE latent representation is learned in an unsupervised manner whereby any notions of interpretability are not explicitly enforced, nor is there any promise that the resulting latent space will adequately encompass human-interpretable features.
Developing techniques to add human-understandable interpretability to the latent space representation is currently an active research area \citep{AdelGhahramaniEtAl2018}.

\begin{figure*}[tb]
    \centering
    \includegraphics[width=\textwidth]{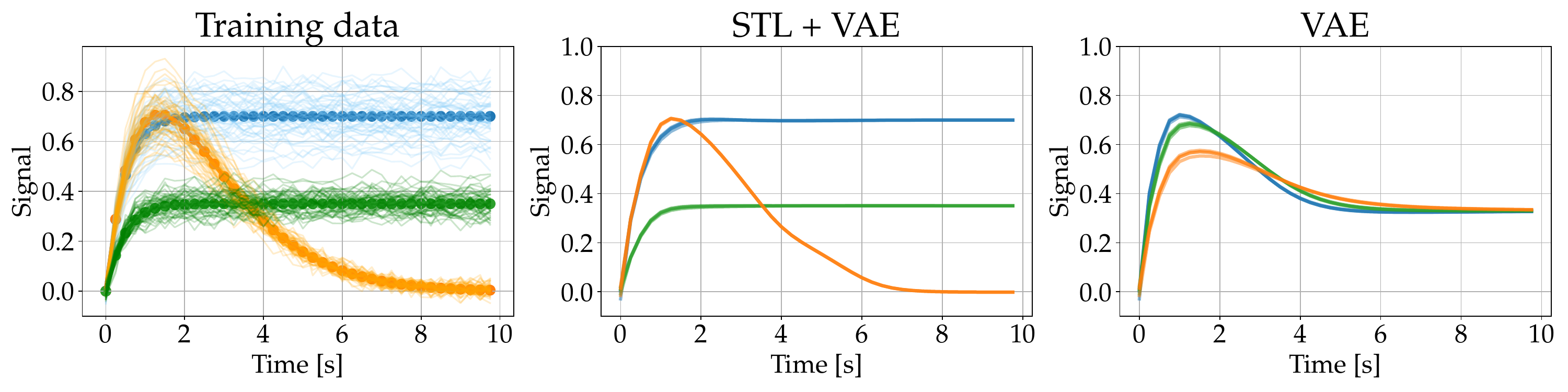}
    \caption{A VAE example leveraging \texttt{stlcg} to provide more structure into the latent space representation. Left: Training data used to train a VAE with a discrete latent space. Middle: Generated outputs from a VAE model leveraging \texttt{stlcg}. Right: Generated outputs from a standard VAE without leveraging \texttt{stlcg}. }
    \label{fig:latent space example}
\end{figure*}

In this example, we demonstrate that we can use \texttt{stlcg} to project expert knowledge about the dataset into the latent space construction, and therefore aid in the interpretability of the latent space representation.
Consider the following dataset illustrated in Figure~\ref{fig:latent space example} (left). The dataset is made up of (noisy) trajectories; Gaussian noise was added onto the trajectories shown in solid color. Notably, the data is \emph{multimodal} as there are three trajectory archetypes contained in the data, each denoted by different colors.
The goal is to learn a VAE model that can generate similar-looking trajectories.
However, training a VAE for multimodal data is generally challenging (e.g., can suffer from mode collapse).

We consider a VAE model with a discrete latent space with dimension $N_z=3$ because, based on our ``expert knowledge'', there are three different types of trajectory modes.
The latent variable $z\in \lbrace 0,1 \rbrace^3$ is a one-hot vector of size three, i.e., $z\in \lbrace [1,0,0]^\top, [0,1,0]^\top, [0,0,1]^\top\rbrace$.
Additionally, we use LSTM networks for the encoder and decoder networks due to the sequential nature of the data.
Based on our ``expert knowledge'', the data satisfies the following STL formula,
\[\phi_c = \square_{[7.5, 9.75]} \left[(x > c-\epsilon) \wedge (s < c + \epsilon)\right],\]
where $c \in \lbrace 0.7, 0.35, 0\rbrace$ and $\epsilon = 0.025$. This STL formula describes the convergence of a signal to a value $c$ within some tolerance $\epsilon$.
Let $C=[0.7, 0.35, 0]^\top$ be a column vector corresponding to the possible parameter values of $\phi_c$ (based on expert knowledge). Then the dot product $z^\top C$ selects the entry of $C$ corresponding to 1's position in $z$.
As such, if $s$ is the output trajectory from the VAE mode, and $z$ is the corresponding latent vector, then we can construct an STL loss, $\mathcal{L}_\mathrm{STL} = \mathrm{ReLU}(-\rho(s, \phi_{z^\top C}))$.
This STL loss enforces each latent vector instantiation to correspond to different STL parameter values and therefore incentives the resulting output to satisfy the associated STL formula.
The overall loss function used to train a VAE model with STL information is,
\[\mathcal{L} = \mathcal{L}_\mathrm{VAE} + \gamma \mathcal{L}_\mathrm{STL},\]
where $\mathcal{L}_\mathrm{VAE}$ is the standard loss function used when training a typical VAE model.

Figure~\ref{fig:latent space example} (middle) illustrates the trajectories generated from a VAE trained with an additional STL loss $(\gamma = 4)$, while Figure~\ref{fig:latent space example} (right) showcases trajectories generated with the same model (i.e., same neural network architecture and size) trained without the STL loss $(\gamma = 0)$.
Note, the hyperparameters using during training (e.g., seed, learning rate, number of iterations, batch size, etc,) were the same across the two models.
We can see that by leveraging expert knowledge about the data and encoding that knowledge via STL formulas, we are able to generate outputs that correctly cover the three distinct trajectory modes.
With more training iterations, it could be possible, though not guaranteed, that the vanilla VAE model would have been able to capture the three distinct trajectory modes. Nonetheless, this example demonstrates that \texttt{stlcg} could potentially be used to accelerate training processes, and enable models to more rapidly converge to a desirable level of performance.

\section{Future Work and Conclusions}\label{sec:conclusion and future work}
In this work, we showed that STL is an attractive language to encode spatio-temporal specifications, either as constraints or a form of inductive bias, for a diverse range of problems studied in the field of robotics.
In particular, we proposed a technique, \texttt{stlcg}, which transcribes STL robustness formulas as computation graphs, and therefore enabling the incorporation of STL specifications in a range of problems that rely on gradient-based solution methods.
We demonstrate through a number of illustrative examples that we are able to infuse logical structure in a diverse range of robotics problems such as motion planning, behavior clustering, and deep neural networks for model fitting, intent prediction, and generative modeling.

We highlight several directions for future work that extend the theory and applications of \texttt{stlcg} as presented in this paper.
The first aims to extend the theory to enable optimization over parameters defining time intervals over which STL temporal operators hold, and to also extend the language to express properties that cannot be expressed by the standard STL language.
The second involves investigating how \texttt{stlcg} can help verify and improve the robustness of learning-based components in safety-critical settings governed by spatio-temporal rules, such as in autonomous driving and urban air-mobility contexts.
The third is to explore more ways to connect supervised structure induced by logic with unsupervised structure learning present in latent spaces models. This connection may help provide interpretability via the lens of temporal logic for neural networks that are typically difficult to analyze.

\section*{Acknowledgements}
This work was supported by the Office of Naval Research (Grant N00014-17-1-2433), NASA University Leadership Initiative (grant \#80NSSC20M0163), and by the Toyota Research Institute (``TRI''). This article solely reflects the opinions and conclusions of its authors and not ONR, NASA, TRI, or any other Toyota entity.

\bibliographystyle{SageH}
\bibliography{../../../../bib/main,../../../../bib/ASL_papers}

\end{document}